**AI and Deep Learning for Terahertz Ultra-Massive MIMO: From Model-Driven Approaches to Foundation Models**

Wentao Yu[a,b], Hengtao He[a], Shenghui Song[a], Jun Zhang[a], Linglong Dai[b,c], Lizhong Zheng[b], Khaled B. Letaief[a,*]

[a] *Department of Electronic and Computer Engineering, The Hong Kong University of Science and Technology, Hong Kong 999077, China*
[b] *Department of Electrical Engineering and Computer Science, Massachusetts Institute of Technology, Cambridge, MA 02139, USA*
[c] *Department of Electronic Engineering, Tsinghua University, Beijing 100084, China*

* Corresponding author.
*E-mail address:* eekhaled@ust.hk (K. B. Letaief).

**ABSTRACT**
This study explored the transformative potential of artificial intelligence (AI) in addressing the challenges posed by terahertz ultra-massive multiple-input multiple-output (UM-MIMO) systems. It begins by outlining the characteristics of terahertz UM-MIMO systems and identifies three primary challenges for transceiver design: computational complexity, modeling difficulty, and measurement limitations. The study posits that AI provides a promising solution to these challenges. Three systematic research roadmaps are proposed for developing AI algorithms tailored to terahertz UM-MIMO systems. The first roadmap, model-driven deep learning (DL), emphasizes the importance of leveraging available domain knowledge and advocates the adoption of AI only to enhance bottleneck modules within an established signal processing or optimization framework. Four essential steps are discussed: algorithmic frameworks, basis algorithms, loss-function design, and neural architecture design. The second roadmap presents channel state information (CSI) foundation models, aimed at unifying the design of different transceiver modules by focusing on their shared foundation, that is, the wireless channel. The training of a single compact foundation model is proposed to estimate the score function of wireless channels, which serve as a versatile prior for designing a wide variety of transceiver modules. Four essential steps are outlined: general frameworks, conditioning, site-specific adaptation, joint design of CSI foundation models, and model-driven DL. The third roadmap aims to explore potential directions for applying pretrained large language models (LLMs) to terahertz UM-MIMO systems. Several application scenarios are envisioned, including LLM-based estimation, optimization, search, network management, and protocol understanding. Finally, the study highlights open problems and future research directions.

## 1. Introduction

### 1.1. Background

Our society is undergoing a digital revolution marked by a drastic increase in both connectivity and throughput [1]. Media transmission has expanded significantly, encompassing images, videos, and, imminently, augmented and virtual reality streams [2,3]. The advent of fifth-generation (5G) mobile communication technology has introduced transformative benefits through the concept of "connected things," while global efforts in sixth-generation (6G) mobile network research and development are expected to usher in a new era of "connected intelligence" with groundbreaking applications and services [4], as shown in Fig. 1. Prominent applications such as the artificial intelligence of things (AIoT), autonomous driving, smart manufacturing, and edge artificial intelligence (AI) are anticipated to play vital roles in 6G and beyond systems. To support these developments, innovative technologies are required to manage the exponential increase in mobile data traffic and its varied applications. Future communication systems must meet demanding requirements in terms of throughput, scalability, latency, and complexity [5,6]. Anticipated performance targets include high data rates of up to 1 terabits per second (Tbps), extremely low end-to-end latencies of less than 100 μs, high spectral efficiency of approximately 100 bits per second per Hertz (bits·$s^{-1}$·$Hz^{-1}$), ultra-wide bands of up to 3 THz, and numbers of connections reaching at least $10^8$ devices per square kilometers. These requirements mandate transformative advances in wireless technology.

Several white papers and technical reports from the International Telecommunication Union (ITU) [7], 5G Americas [8], and China's IMT-2030 (6G) promotion group [9] have emphasized the importance of exploring unexplored higher frequency bands for 6G and beyond systems. Among these, the terahertz band, that is, the spectrum spanning 100 GHz to 10 THz, remains largely underutilized and presents opportunities to fulfill the ever-increasing demand for wireless links [10]. The US Federal Communications Commission (FCC) has already allocated 95 GHz–3 THz spectrum for 6G, positioning the United States as a leader in the 6G race [11]. Furthermore, the Institute of Electrical and Electronics Engineers (IEEE) 802.15.3d standard has initiated preliminary standardization for terahertz-band communications [12]. Along with the quest for higher frequency bands, a natural trend exists toward deploying more antennas at base stations (BSs). These developments have evolved from small-scale multiple-input multiple-output (MIMO) systems with only a few antennas in third-generation (3G) to large MIMO systems in fourth-generation (4G)/5G systems [13].

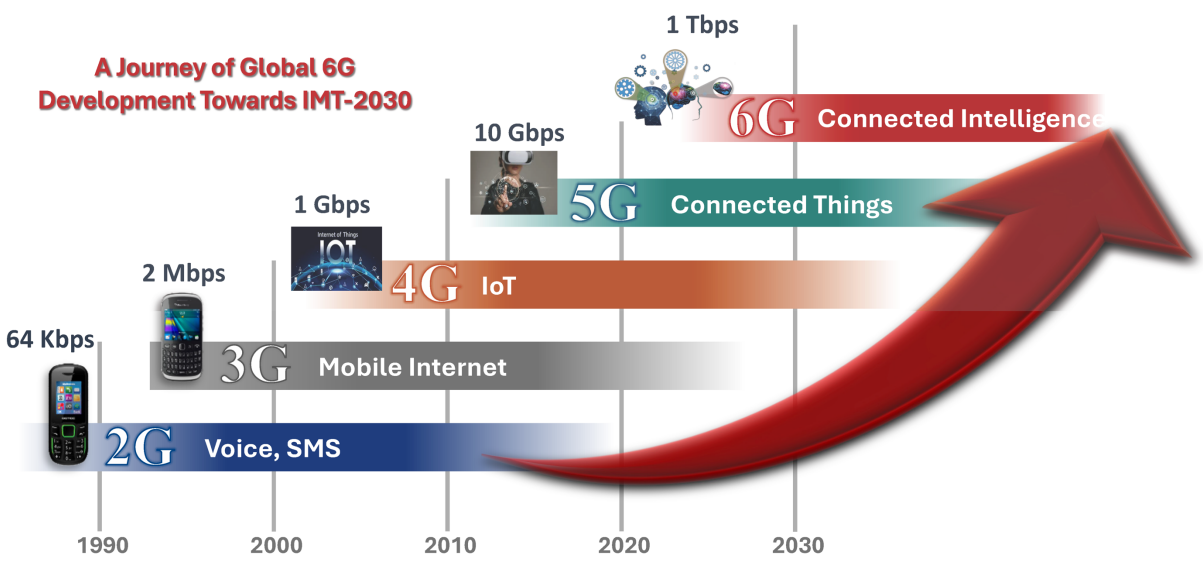

**Fig. 1.** Journey of global wireless development toward IMT-2030. Tbps: terabits per second;Gbpsl: gigabits per second; Mbps: megabits per second; kbps: kilobits per second; IoT: Internet of Things; SMS: short message service; 2G: second-generation.

*1.2. UM-MIMO systems in the terahertz band*

Looking ahead to future wireless networks, ultra-massive MIMO (UM-MIMO) arrays with more than 1024 antennas are expected to be deployed in the terahertz frequency band [14–16]. This transformative technology can address significant path loss and molecular absorption loss by employing highly directional beamforming to enhance coverage [17] and leveraging the abundant bandwidth to achieve high spectral and energy efficiencies [18]. Additionally, terahertz UM-MIMO offers higher localization accuracy with reduced transmission power and a smaller footprint compared to its millimeter-wave (mmWave) counterparts [19], while supporting the integration of sensing and communication functionalities [20,21]. Another key application is in nanocommunications, such as on-chip communications and in-body networks, which benefit from the compact size of terahertz arrays built using nano- and meta-materials [22,23]. In addition, line-of-sight (LoS) terahertz UM-MIMO arrays have the potential to replace copper or fiber point-to-point links in data center networks and provide high-capacity links for aerial and space networks, enabled by spherical-wave spatial multiplexing in the near-field of the array [24,25]. terahertz-band communications are also poised to improve connectivity and physical-layer security in 6G and beyond systems [26].

Nevertheless, the study of terahertz UM-MIMO systems is still in its early stages and several unique challenges must be addressed [27]. These challenges can be summarized by three "hard-to" problems from a signal processing perspective. First, the massive system scale and short coherence times in terahertz UM-MIMO systems make many traditional model-based designs too complex, resulting in a "hard-to-compute" problem. This necessitates low-complexity, real-time algorithms capable of handling high-dimensional signal processing and optimization tasks efficiently. Second, the complex channel characteristics in the terahertz band give rise to several new phenomena, such as hybrid far- and near-field effects (also known as the hybrid-field effect [28,29]), spatial non-stationary effect [30,31], and wideband beam squint effect [32]. Together, these phenomena present a "hard-to-model" challenge in transceiver design, as classical optimization and analysis methods often rely on precise system and channel modeling. Overcoming this requires innovative approaches that can learn from and adapt to complex environments without relying on analytical models. Third, there is the "hard-to-measure" problem, specifically referring to channel measurement. Although "hard-to-measure" may also imply difficulties in evaluating the performance of AI-based communication systems using traditional metrics, in this study, the phrase refers solely to the challenges in channel measurement. The array-of-subarray (AoSA) architecture commonly used in terahertz UM-MIMO systems results in far fewer radio frequency (RF) chains than the number of antennas [33,34]. Combined with hardware impairments [35–37], this architecture produces incomplete and corrupted measurements during channel estimation, significantly complicating the acquisition of accurate channel state information (CSI) [38]. To overcome these limitations and to enhance the performance of CSI acquisition and channel-dependent tasks, novel solutions should be studied to leverage historical data and overcome the drawback of corrupted measurements.

Notably, UM-MIMO has also been investigated in the literature using other terminologies, such as extremely/extra large-scale MIMO (XL-MIMO) systems [31,39–41], extremely large aperture array (ELAA) [42,43], and gigantic MIMO (gMIMO) [44] for different frequency bands. In this study, given the focus on terahertz communications, we adhere to the terminology terahertz UM-MIMO, originally proposed in Ref. [15], as it was the first to introduce this concept specifically for the terahertz band.

### *1.3. AI for communications*

Since 2017 [45], AI, particularly deep learning (DL), has regained attention in the wireless communications community and has become an indispensable tool for the design and optimization of large-scale multi-antenna systems [46,47]. The applications of AI can be categorized according to the three "hard-to" challenges discussed previously. Firstly, one potential application of AI is to exploit its capability to facilitate the solution of "hard-to-compute" problems, those traditionally considered intractable owing to challenges such as high-dimensionality, high-cardinality, or non-convexity [48]. AI models can be trained to directly and implicitly approximate the desired solution or function by replacing bottleneck modules in established optimization or signal processing algorithms with learnable components [49]. These two paradigms are called data-driven and model-driven approaches in wireless communication literature [50,51]. In addition, AI is also helpful in tackling "hard-to-model" problems by learning complex and non-linear relationships from data, eliminating the need for accurate analytical models. This has been particularly successful in high-dimensional problems at the physical layer and model-free problems at the medium access control (MAC) layer and above [46,47]. Finally, "hard-to-measure" problems may be addressed using generative AI [52] and foundation models. These approaches learn channel distributions from historical data and generate high-fidelity synthetic channels for data augmentation, thereby enhancing channel-dependent tasks. Furthermore, they can serve as an open-ended prior for solving inverse problems such as channel estimation, prediction, and tracking, in the physical layer using incomplete and corrupt measurements [52].

After nearly eight years of exploring AI for communication, a few shared perspectives have gradually emerged in this field. The first is that AI is especially valuable when analytical models fail to deliver optimal solutions [46,48,53,54], whether due to excessive complexity (i.e., the "hard-to-compute" problem) or the lack of an accurate model (i.e., the "hard-to-model" problem). This understanding has given rise to the model-driven DL paradigm, where AI augments traditional model-based methods by replacing their bottleneck components with learnable modules, thereby integrating expert knowledge with learning capabilities [50,55]. The second perspective is that data will probably be a major bottleneck for training AI models in wireless communications [53,56–58]. Large-scale data collection and channel measurement are required to characterize site-specific channel and user distributions, which are both expensive and time-consuming (i.e., the "hard-to-measure" problem). The third perspective, based on our own understanding, is that wireless transceivers share a common foundation—the wireless channel. This raises the question of whether a unified foundation model can be developed to serve various transceiver modules, rather than designing separate AI models for each problem. A unified model can reduce both training and deployment costs. These last two perspectives inspired us to propose a new concept: the CSI foundation model. Based on the available data, we aim to train a single, compact generative AI model capable of capturing site-specific channel characteristics for data augmentation and serving as a versatile prior for designing a wide range of transceiver modules.

### *1.4. Contributions and organization*

Although terahertz UM-MIMO is envisioned as a promising candidate for 6G and beyond wireless systems [59], research in this field is still in its early stages, with significant potential for future development. As the field continues to evolve, researchers interested in terahertz UM-MIMO systems may wish to apply AI to solve various emerging challenges but lack guidance on where to begin. Conversely, experts in AI for communications may not be familiar with the distinct features of terahertz systems, limiting their ability to contribute. This study aims to bridge these gaps by introducing the key system and channel characteristics of terahertz UM-MIMO, identifying the associated challenges, and illustrating how AI can be leveraged to address them. We propose two systematic research roadmaps—model-driven DL and CSI foundation models—and provide a step-by-step guide for navigating each. We highlight key frameworks and techniques essential for developing AI-based solutions for terahertz UM-MIMO systems. Through this study, we seek to foster collaboration between experts in AI and terahertz communications to advance this exciting interdisciplinary field [60,61].

The organization of this paper is illustrated in Fig. 2. Section 1 provides the introduction. Section 2 discusses the preliminaries of terahertz UM-MIMO communications and highlights key system and channel characteristics. Section 3 examines how the three "hard-to" challenges manifest in terahertz UM-MIMO systems and explain why AI is well-suited to tackle these challenges. Section 4 presents research roadmaps for developing model-driven DL and CSI foundation models for terahertz UM-MIMO systems, illustrating their design principles, key components, and representative case studies. Finally, Section 5 concludes the paper.

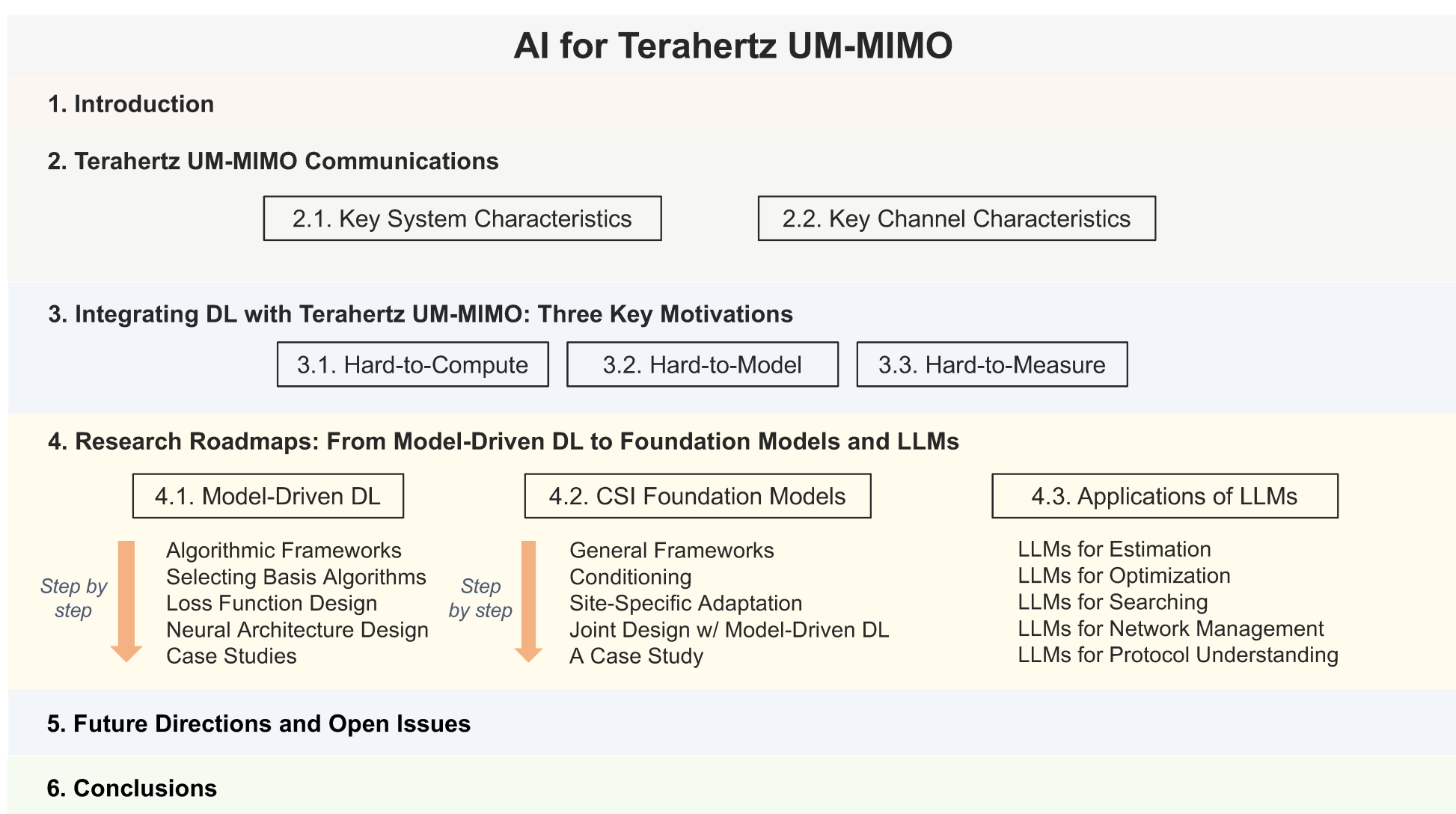

**Fig. 2.** Organization of this paper. LLMs: large language models.

## 2. Terahertz UM-MIMO communications

In this section, we introduce the key features and channel models of terahertz UM-MIMO systems and explain the challenges they present in transceiver design. Because an overview paper [18] already offers a comprehensive introduction to the mathematical models of terahertz UM-MIMO, we avoid redundancy and direct readers to Section 2 in Ref. [18] for more details. Open-source codes for terahertz UM-MIMO systems are available in Ref. [38].

### *2.1. Key system characteristics*

Terahertz-band waves experience severe propagation path loss due to molecular absorption at terahertz frequencies [62,63]. UM-MIMO antenna systems, comprising thousands of antennas, have recently emerged as promising solutions to overcome this limitation [15,16]. The small wavelength in the terahertz band allows the deployment of a large number of antennas with a small footprint. Consequently, super-narrow beams are formed, which help mitigate path losses and expand communication range. The practical implementation of terahertz UM-MIMO has become feasible with the development of new plasmonic materials, such as graphene and nanomaterials, that enable the construction of nanoantennas and transceivers for terahertz band communication [23].

#### *2.1.1. AoSA architecture*

Terahertz UM-MIMO systems face constraints owing to the high-complexity and power-consuming demands of terahertz hardware. These costly components preclude traditional designs used in lower-frequency bands. A fully digital architecture with one RF chain per antenna is infeasible because of its high cost. Similarly, the fully connected hybrid analog-digital architecture for mmWave bands [33,64], where a limited number of RF chains drive the entire antenna array, is also inefficient in the terahertz regime due to transmit power and circuit feeding limitations. In this setup, each RF chain must be connected to all antennas via phase shifters to perform analog beamforming or combining [34]. This requires a significant number of phase shifters in the terahertz band, making it expensive to implement.

Consequently, the prevailing choice for terahertz UM-MIMO arrays is a simplified architecture called AoSA [18,34]. In this architecture, a UM-MIMO array is grouped into various nonoverlapping subarrays (SAs), with each RF chain connecting to and powering only its corresponding SA, as illustrated in Fig. 3(a) [29]. This is similar to the partially connected architecture used in mmWave bands [33,65]. AoSA significantly reduces the number of required phase shifters in the analog beamformer (or combiner) and lowers energy consumption.

Fig. 3(b) [29] illustrates a magnified view of a part of the terahertz UM-MIMO array. The spacing between a pair of adjacent antenna elements (AEs) is denoted as $d_{\mathrm{a}}$, and that between adjacent SAs is represented by $d_{\mathrm{sub}} \triangleq \omega d_a$ $(\omega \gg 1)$, where $\omega$ is a constant. Here, $d_{\mathrm{a}}$ is typically small owing to the small wavelength in the terahertz band. Conversely, the SAs should be separated by a much larger distance (i.e., $\omega \gg 1$), because closely integrating too many AEs can create difficulties in the control and cooling of circuits [18,66]. As a result, terahertz UM-MIMO arrays are often nonuniform, differing from conventional uniform arrays. This non-uniformity necessitates specialized considerations in algorithm design [38,67].

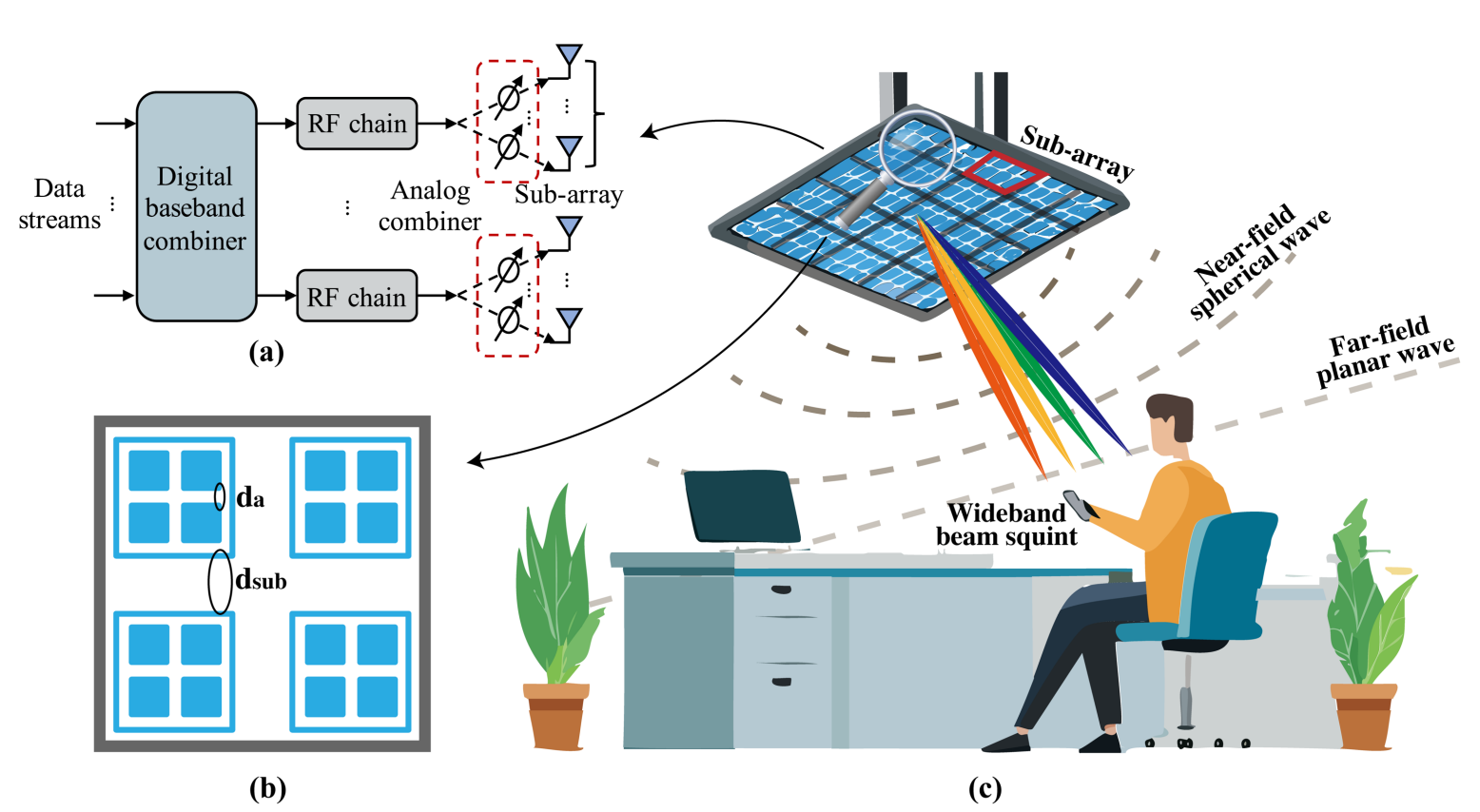

**Fig. 3.** (a) AoSA architecture partitions the terahertz UM-MIMO array into various SAs, each driven by a dedicated RF chain. Each SA is connected with its designated RF chain via a phase shifter network. (b) The zoom-in view of a part of the terahertz UM-MIMO array, in which each SA is labeled by a blue frame, while each AE is denoted by a blue square. (c) A typical terahertz UM-MIMO system in an indoor office. The wavefront from the terahertz UM-MIMO array varies with the distance between the transmitter and the receiver: It is spherical in the near-field region and (approximately) planar in the far-field region. Due to the multi-path propagation, the terahertz UM-MIMO channel is typically a mixture of far- and near-field components, called the hybrid-field effect [29]. The wideband beam squint effect will also be observed in terahertz UM-MIMO, where the beams at different subcarriers tend to split into different physical directions. $d_{\mathrm{a}}$: the spacing between a pair of adjacent antenna elements; $d_{\mathrm{sub}}$: the spacing between adjacent SAs.

*2.1.2. Beam squint effect*

The AoSA architecture illustrated in Fig. 3(a) [29] is used for narrowband terahertz UM-MIMO systems. However, in wideband terahertz systems, a new challenge known as beam squint arises, requiring modifications to the original narrowband architecture [32]. In wideband terahertz UM-MIMO systems, an ultrawide bandwidth and a large number of antennas can result in a non-negligible propagation delay across the antenna array, which may exceed the sampling period. This leads to variations in the angles of departure (AoDs) or arrival (AoAs) across different subcarriers during transmission or reception, making the array gain frequency selective. Given this characteristic, if frequency-flat phase shifters are adopted, the resultant beams will be dispersed and will point towards various angles, as shown in Fig. 3(c) [29]. To address the beam squint effect, various studies have adopted true-time-delay (TTD) modules to compensate for angle variations [68,69], such as delay-phase precoding [70] and joint phase–time arrays [71]. TTD modules are typically deployed between each RF chain and the associated phase shifters. The number of TTD modules is usually smaller than that of the phase shifters because of their relatively higher costs [68]. In addition, the beam-squint effect can be exploited to accelerate wideband terahertz beam tracking by controlling the degree of the squint to scan multiple angular directions simultaneously [72,73].

*2.2. Key channel characteristics*

Measurement campaigns are currently underway worldwide to further the understanding of terahertz propagation characteristics. For comprehensive surveys, readers are referred to Refs. [74,75]. As this study focuses on physical layer signal processing in terahertz UM-MIMO systems, we summarize only the most relevant channel features here.

*2.2.1. Hybrid-field effect*

Near- and far-field phenomena are crucial at terahertz frequencies [76]. In the far-field region, the wavefront can be assumed to be approximately planar. However, when signals originate in the near-field region, this assumption no longer holds [77]. An exactly spherical wavefront model must be considered when modeling the terahertz UM-MIMO channels with near-field multipath components, as illustrated in Fig. 3(c) [29].

The boundary between the far-field and the (radiating) near-field regions is called the Rayleigh distance, defined by $\frac{2D^2}{\lambda_{\mathrm{c}}}$, where $D$ is the array aperture, and $\lambda_{\mathrm{c}}$ denotes the carrier wavelength. While the Rayleigh distance appears to increase linearly with frequency for a fixed aperture, this relationship can be misleading. In practice, the array aperture $D$ is also a function of the carrier wavelength, as the AE spacing is typically set to half the wavelength. Consider a planar AoSA with $\sqrt{S}\times\sqrt{S}$ SAs with $d_{\mathrm{sub}} \triangleq \omega d_{\mathrm{a}}$ $(\omega \gg 1)$, where each SA is a uniform planar array with $\sqrt{\bar{S}}\times\sqrt{\bar{S}}$ AEs, where $S$ and $\bar{S}$ are both constants. In this case, the Rayleigh distance is given by $\left\{\sqrt{S}\left(\sqrt{\bar{S}}-1\right)+\left(\sqrt{S}-1\right)\omega\right\}^2 \lambda_{\mathrm{c}}$ [38]. This shows that the boundary between the far- and near-field regions is jointly determined by the carrier wavelength and array geometry and should be analyzed on a case-by-case basis.

In a general scenario, sources and scatterers can be positioned in both the far- and near-field regions of a UM-MIMO array. Consequently, different multipath components can arrive at the array as spherical or planar wavefronts. Hence, terahertz UM-MIMO channels typically consist of a dynamic mixture of both, which have been identified in the context of channel estimation as the hybrid-field effect [29,38,78]. In the literature, this phenomenon is referred to as the cross-field effect [28,79], although both terms convey a similar concept.

Spherical wavefronts provide important benefits such as enhanced spatial multiplexing, high-precision localization, and transverse velocity sensing. However, they complicate the representation, acquisition, and exploitation of channels, particularly when mixed with far-field planar wavefronts, due to the hybrid-field effect.

*2.2.2. Multipath components*

Terahertz channels are typically sparse, with a limited number of resolvable paths, owing to high scattering and diffraction losses in the terahertz band. For example, Yan et al. [80] reported that at 300 GHz, the number of multipaths was only five for a $256\times256$ UM-MIMO array, which was 32.5% less than its counterpart at 60 GHz. Intuitively, the terahertz-band channel exhibits a higher *K*-factor than the lower-frequency bands owing to greater reflection and diffraction losses [75]. Although we emphasize LoS-dominant propagation, multipath components should still be considered, particularly in indoor terahertz communications. Existing measurement campaigns are mostly conducted within the sub-terahertz band (i.e., 100–300 GHz). Further investigation is needed for the full terahertz band.

*2.2.3. Molecular absorption effect*

When terahertz electromagnetic waves propagate through a nonvacuum medium, they can trigger resonances in certain molecules along their paths, resulting in notable energy loss at certain frequencies. From a communications perspective, this causes strong frequency-selective channel gains at different subcarriers in a wideband terahertz system. This phenomenon is known as the molecular absorption effect. The absorption strength can vary with environmental conditions, such as humidity and temperature, and depends on the propagation distance. A detailed frequency profile of these absorption peaks can be found in Ref. [63]. These peaks divide the terahertz band into multiple narrow spectral windows with relatively low absorption. Therefore, terahertz communication should only be placed within these spectral windows [63]. Conversely, molecular absorption loss can be leveraged for designing distance-adaptive absorption peak modulation to improve the covertness of terahertz communications [81].

*2.2.4. Spatial non-stationary effect*

The spatial non-stationary effect occurs when the terminals or scattering clusters are visible only from a portion of the UM-MIMO array [31]. This phenomenon is more prevalent in linear arrays, where the array aperture is larger than in planar arrays with the same number of AEs. Additionally, the lower-frequency end of the terahertz spectrum is more susceptible to spatial non-stationarity in the sub-terahertz band, where the larger wavelengths result in increased AE and SA spacings, and hence larger array apertures. In contrast, planar arrays operating at higher terahertz frequencies exhibit less pronounced spatial non-stationarity owing to their smaller apertures. Therefore, a novel three-dimensional (3D) non-stationary geometry-based stochastic model has been proposed for terahertz UM-MIMO systems [30].

## 3. Integrating DL with terahertz UM-MIMO: Three key motivations

We believe that DL is especially effective in tackling three key challenges in wireless communications—that is, “hard-to-compute,” “hard-to-model,” and “hard-to-measure.” In this section, we discuss the manifestation of these challenges in terahertz UM-MIMO systems, this explains the motivation for integrating DL with terahertz UM-MIMO systems.

### 3.1. *"Hard-to-compute"problems*

In terahertz UM-MIMO networks, the system scales drastically in terms of network density, number of antennas, supportable users, and system bandwidth. This leads to high-dimensional signal processing and optimization problems and presents challenges in terms of computational complexity. Additionally, the channel coherence time in terahertz UM-MIMO networks is extremely short, resulting in rapid channel fluctuations. Conventional statistical and optimization-based approaches involving computationally intensive operations, such as singular value decomposition (SVD), bisection search, and matrix inversion, may struggle to meet latency requirements. These challenges highlight the importance of low-complexity methods.

Unlike traditional methods, DL excels at making fast approximations to avoid heavy computation [48]. It can learn intricate patterns and solve large-scale signal processing and optimization problems approximately in real time. Traditional algorithms treat every problem instance as completely new and solve each from scratch; however, DL can recognize similarities between incoming problem instances and find a shortcut [49]. When trained and fine-tuned in a site-specific manner, DL can efficiently learn the distribution of "problem instances" unique to that environment, allowing it to outperform general algorithms in terms of performance and efficiency [82]. In addition, DL models can operate in parallel across various wireless resource domains. For example, in wideband terahertz UM-MIMO systems, different SAs and subcarriers can be computed simultaneously. The industry is also actively developing AI-native radio access network (AI-RAN) algorithms that leverage the powerful parallel computation capabilities of graphical processing units (GPUs) [83].

### 3.2. *"Hard-to-model" problems*

Modeling capability is also a key motivation for applying DL. The success of classical mathematical tools such as optimization and analysis highly depends on the accuracy of system and channel models. Because the network architecture, communication environment, and wireless channels in terahertz UM-MIMO systems are heterogeneous, complex, and non-linear, analytical models struggle to capture them precisely. In terms of network architecture, the limited coverage of terahertz systems necessitates further densification, which can cause complex interference issues. Additionally, narrow beams increase the possibility of misalignment. In terms of the communication environment, terahertz waves are highly susceptible to spatial non-stationarity, signal blockage, and frequency-selective molecular absorption loss. Real-world networks are significantly more complex than simplified system models due to these combined factors. Most importantly, as mentioned in Section 2.2, terahertz UM-MIMO channels may consist of a dynamic mixture of planar-wave and spherical-wave multipath components due to the hybrid-field effect [28,38]. Consequently, the widely used angle-domain sparsity properties or simplified prior distributions in traditional far-field systems are no longer applicable. This makes it difficult to estimate and track wireless channels [84,85].

### 3.3. *"Hard-to-measure" problems*

As discussed previously, terahertz UM-MIMO systems often utilize the AoSA architecture to enhance energy efficiency and reduce costs. Such architectures have far fewer RF chains than antennas [34]. The signals received in each timeslot provide only an incomplete measurement of the channels. Moreover, the presence of frequency-selective noise, hardware distortions, and impairments adds another layer of complexity to the measurement process [36,37,86]. Incomplete measurements and impairments significantly complicate the estimation and tracking of channels and environmental dynamics [87–90].

Wireless channels are pivotal in transceiver design. Difficulties in acquiring CSI can lead to performance degradation in many downstream tasks that rely on channel statistics. To address these problems, we can pretrain generative AI models offline using site-specific historical data to serve as priors, compensating for missing information due to incomplete measurements [57,91,92]. Such AI-based priors, when incorporated with appropriate algorithmic frameworks, can solve versatile tasks including channel estimation [93,94], data detection [95,96], CSI compression and feedback [97], and RF source separation [98]. In addition, they can be directly sampled to perform data augmentation and generate synthetic channel samples to further facilitate the design of transceiver modules.

## 4. Research roadmaps: From model-driven DL to foundation models and LLMs

In this section, we present three systematic roadmaps for developing AI-enabled solutions to address the key challenges posed by terahertz UM-MIMO systems. These roadmaps include model-driven DL models, CSI foundation models, and large language model (LLM) applications. Our goal is to inspire readers to follow these roadmaps to design AI-enabled solutions for their own research in terahertz UM-MIMO systems. We outline the three roadmaps, consisting of essential steps and case studies, in Fig. 4. By following these steps, one should be able to develop effective AI-enabled algorithms for various problems in terahertz UM-MIMO. Additionally, we discuss how the roadmaps intersect and how they can be seamlessly integrated. Because we aim to structure this paper as a roadmap, we refer readers to the existing literature for in-depth discussions on these topics and provide detailed explanations only for those lacking adequate guidance. Although these roadmaps are primarily designed for the characteristics of terahertz UM-MIMO systems, most of them are also backward-compatible with traditional MIMO and massive MIMO systems in the sub-6 GHz, upper-mid, and mmWave bands. This is because these systems typically operate on a smaller scale and exhibit simpler channel characteristics.

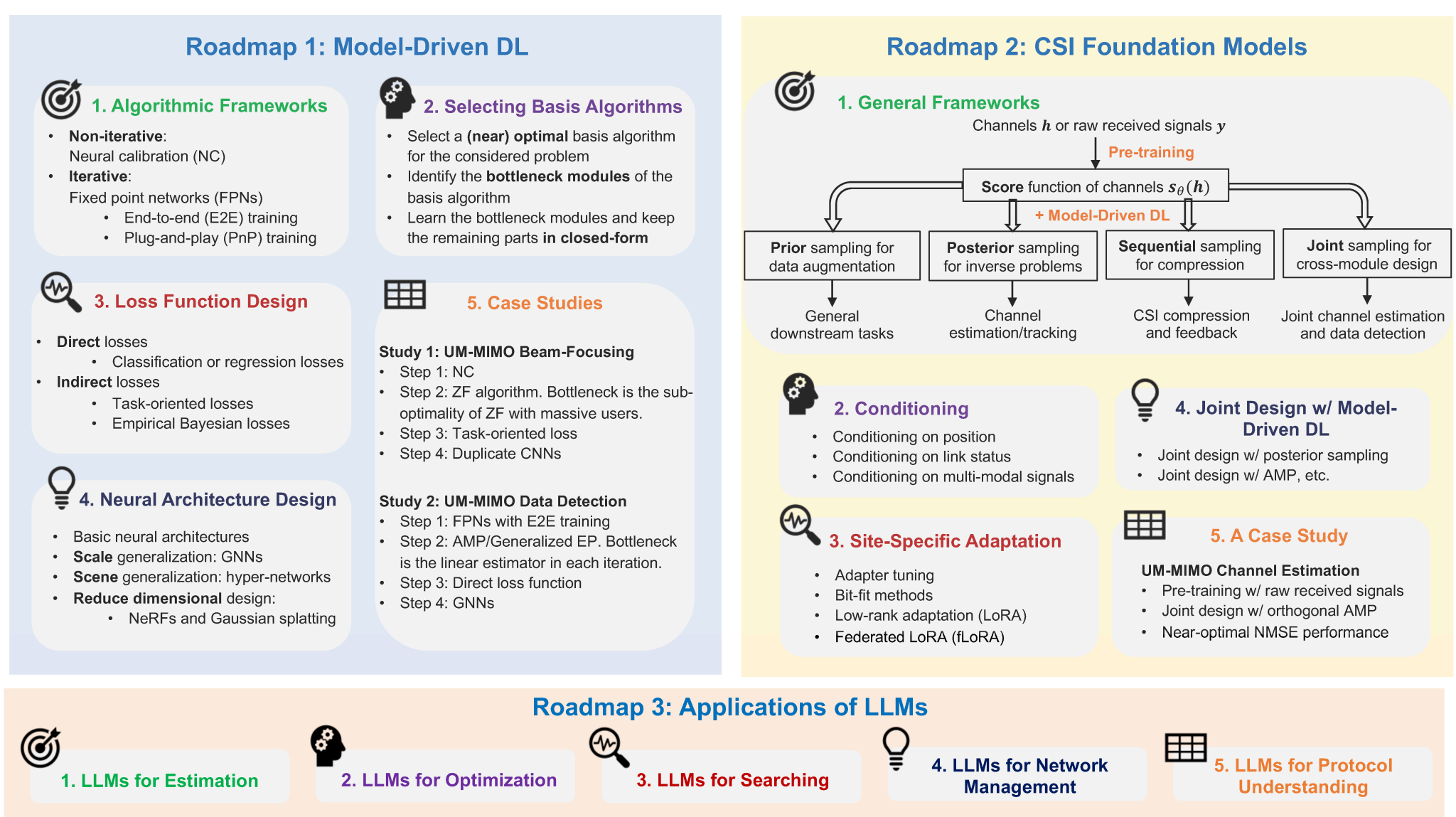

**Fig. 4.** Proposed research roadmaps of AI for terahertz UM-MIMO: from model-driven DL to CSI foundation models and LLMs. GNN: graph neural network; NeRF: neural radiance fields; ZF: zero-forcing; CNN: convolutional neural network; EP: expectation propagation; w/: with; AMP: approximate message passing; NMSE: normalized mean square error.

For ease of reading, a summary of the AI/DL methods and frameworks is provided in Table 1 [38,91,92,94,99–114].

**Table 1**

Summary of AI/DL methods and frameworks discussed in this paper.

| Roadmaps | AI/DL methods and frameworks | Key problems solved | Ref. |
| --- | --- | --- | --- |
| Roadmap 1: model-driven DL | Fixed point networks | Providing a general framework for DL-based iterative algorithms | [38] |
| | Neural calibration | Providing a general framework for DL-based non-iterative algorithms | [99] |
| | Task-oriented losses | Avoiding the computationally challenging label generation process | [100] |
| | Empirical Bayesian losses | Enabling unsupervised learning and adaptation in unknown environments | [101] |
| | Graph neural networks | Enabling generalization to mismatched numbers of users and antennas | [102] |
| | Hyper-networks | Enabling generalization in dynamic wireless environments | [103] |
| | Learning on the transform domain | Reducing sample complexity via low-dimensional representation | [104] |
| | Neural RF radiance field and Gaussian splatting | Learning low-complexity location-to-channel mapping | [105] |
| Roadmap 2: CSI foundation models | Denoising score matching | Training CSI foundation models using channel data or raw received signals | [94] |
| | Prior sampling | Data augmentation, generating synthetic channels via CSI foundation models | [91] |
| | Posterior sampling | Bayesian inference with prior information from CSI foundation models | [106] |
| | Sequential sampling | CSI compression and decompression based on CSI foundation models | [107] |
| | Joint sampling | Bayesian inference based on CSI foundation models and symbol distributions | [108] |
| | Conditioning | Integrating side information into CSI foundation models | [92] |
| | Low-rank adaptation | Efficient fine-tuning of CSI foundation models via low-rank adaptation | [109] |
| Roadmap 3: applications of LLMs | LLMs for estimation | Serving as backbone or hyper-network for parameter estimation | [110] |
| | LLMs for optimization | Solving optimization problems described by natural languages | [111] |
| | LLMs for searching | Solving high-complexity combinatorial problems offline | [112] |
| | LLMs for network management | Autonomous network management via LLM agents | [113] |
| | LLMs for protocol understanding | Extracting key insights from protocols via telecom-specific LLMs | [114] |

### *4.1. Roadmap 1: Model-driven DL*

#### *4.1.1. Overview*

DL algorithms for communications can be categorized into two different paradigms: model-driven DL and (fully) data-driven DL. The data-driven paradigm trains neural networks, such as multilayer perceptrons (MLPs), to map system parameters directly to the desired outputs without relying on domain knowledge or existing algorithms. This approach leverages the universal approximation capabilities of neural networks to learn solely from data [115]. Early attempts at applying DL to communications mostly followed this paradigm. For example, an MLP was trained to map the received pilot and data blocks to the detected data symbols [116]. In another example [117], an MLP was trained to learn the direct mapping between the inputs and outputs of an optimization algorithm for interference management.

However, many physical-layer problems involve well-defined system models and established algorithms that already offer efficient solutions. A data-driven paradigm may overlook valuable domain knowledge of the system. While neural networks can, in principle, learn this knowledge from large amounts of data, such an approach is not sample-efficient and requires substantial training data—akin to reinventing the wheel. As training data in wireless systems are often difficult to collect, it is essential to improve the performance of DL tools using the limited available data. Additionally, unlike traditional signal processing and optimization algorithms that can directly adapt to different system parameters, such as the number of antennas, users, and signal-to-noise ratios (SNRs), data-driven DL models often lack generality and flexibility. The outputs of data-driven DL methods may not conform to the physical constraints defined by wireless system models. Finally, data-driven DL approaches often cannot provide the same level of theoretical guarantees as traditional algorithms.

Model-driven DL is a framework that integrates system domain knowledge into the design of DL models. In particular, model-driven DL replaces bottleneck modules within established signal processing or optimization algorithms with neural network-based learnable components. Often, we recognize that the overall algorithmic framework is optimal; however, its practical implementation is hindered by specific bottlenecks. These bottlenecks may stem from “hard-to-compute” problems, where certain parts of the algorithm are computationally intensive; “hard-to-model” problems, where no analytical solution exists for some components; or “hard-to-measure” problems, where it is difficult to obtain sufficient channel data for training. The earliest work on model-driven DL in communications dates to Ref. [118], where a denoising convolutional neural network (DnCNN) was incorporated as a non-linear denoiser within the framework of the approximate message passing (AMP) algorithm. Following this pioneering study, numerous works have been conducted based on model-driven principles [119].

In the following section, we break down the procedure for designing model-driven DL algorithms into four steps, as shown in Fig. 4: determining algorithmic frameworks, selecting basis algorithms, loss function design, and neural architecture design. By designing the neural architecture and loss function of the DL components, we enforce them to generalize across system scales and dynamic environments and reduce the requirement for clean channel data and optimal labels in the training process. In addition, when combined with CSI foundation models, the DL components in model-driven DL can be shared, which significantly reduces deployment costs. We guide readers through each of these steps and present two representative case studies that apply model-driven DL to terahertz UM-MIMO systems.

#### *4.1.2. Determining algorithmic frameworks*

This is the first step in designing model-driven DL algorithms. The problem at hand is analyzed, and the decision to employ iterative or noniterative algorithms is taken depending on the computational and memory budgets. We propose two general algorithmic frameworks that are scalable, efficient, and theoretically guaranteed. For iterative algorithms, we advocate the fixed point networks (FPNs) framework [38,120] shown in Fig. 5(a) [85], whereas for noniterative algorithms, we discuss the neural calibration (NC) framework in Fig. 5(b) [99].

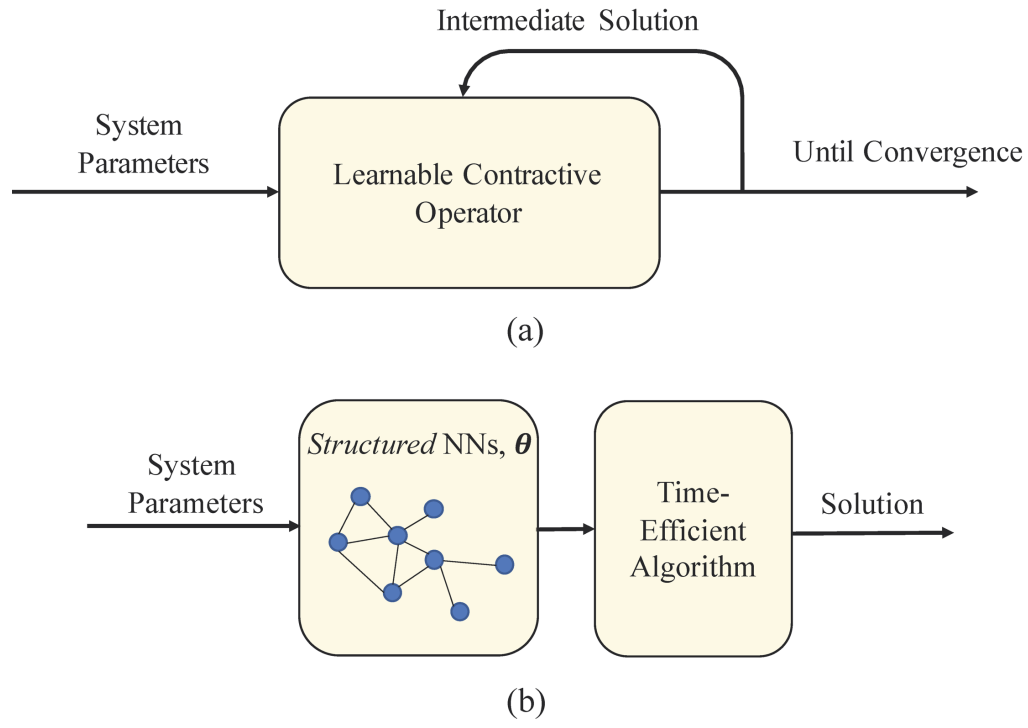

**Fig. 5.** Algorithmic frameworks in model-driven DL [85]. (a) FPNs for iterative algorithms; (b) NC for non-iterative algorithms. NNs: neural networks.

The idea of FPNs was inspired by the fact that many widely used iterative algorithms in wireless communications, such as AMP and proximal algorithms, can be regarded as fixed-point iterations of a contractive operator [121]. A contractive operator is characterized by a Lipschitz constant of less than 1, ensuring linear convergence to a unique fixed point through a fixed-point iteration. To design FPNs, we first identify the contractive operator associated with the target iterative algorithm. We then replace the bottleneck modules with a neural network and construct a learnable operator. The learnable operator is trained to be contractive, with a fixed point that closely approximates the desired solution. This framework is illustrated in Fig. 5(a) [85].

The training process for FPNs aims to achieve two key objectives. First, the learnable mapping must be contractive to ensure fast and monotonic convergence, which can be achieved using regularization techniques [29,122]. The Lipschitz constants of compositions of different mappings equal the product of their respective Lipschitz constants. Thus, we can calculate the necessary Lipschitz constant for the neural network and use regularization techniques to ensure that the learnable mapping is contractive. Second, the fixed point of the learnable operator must closely approximate the desired solution. Training can be conducted in two distinct modes: end-to-end (E2E) and plug-and-play (PnP).

- **E2E mode:** In this approach, we train the learnable components of the contractive operator by directly optimizing the overall performance of the iterative algorithm. For instance, Yu et al. [38] trained a learnable operator to optimize the final estimation accuracy.
- **PnP mode:** In this mode, the training process begins by identifying the specific role that the learnable module assumes within the iterative algorithm, as well as the Lipschitz constant required for convergence. For instance, if a minimum mean-square-error (MMSE) denoiser is required in an iterative algorithm, we can independently train a module that fulfills this role. Once trained, this module can be "plugged" into certain iterative algorithms to enhance its performance, as demonstrated previously in Ref. [101]. One advantage of the PnP scheme is that the learnable module can be separated from a specific algorithm; hence, it can be reused in various algorithms.

The contractive-mapping property offers several advantages for FPNs, including but not limited to the following:

- **Convergence:** The Banach fixed-point theorem guarantees that each iteration brings the solution closer to a fixed point, thereby ensuring monotonic convergence.
- **Adaptive performance-complexity tradeoff:** Owing to monotonic convergence, more iterations correspond to a closer distance to the fixed point and better performance. The number of iterations can be adaptively controlled according to the computational budget.
- **Scalability and low complexity:** Based on the implicit function theorem, the gradient in E2E training can be computed with constant complexity regardless of the number of iterations [122]. The computation relies only on a fixed point and does not require storage of intermediate states during the iteration. In the PnP mode, training is independent of the iterative algorithms and avoids the complexity caused by iterations. Its low complexity makes it favorable for large-scale systems.

Deep unfolding/unrolling networks (DUNs) discussed in literature refer to concepts similar to FPNs [100,123]. Both DUNs and FPNs aim to replace bottleneck modules in existing iterative algorithms with neural networks. The difference lies in the training process; DUNs do not impose contractive constraints. Consequently, they lack several advantages that are unique to FPNs. For instance, DUNs generally do not guarantee convergence, their computational complexity is not adaptive, and their training process requires the storage of all intermediate states, resulting in high computational and memory demands [85]. Most DUNs can be transformed into FPNs by enforcing contractive constraints. We recommend FPNs over DUNs because they are easy to implement and offer several advantages.

Subsequently, we introduce the NC framework for non-iterative algorithms [99]. This can be considered a degenerate version of the FPNs with only one iteration. As shown in Fig. 5(b) [85], the NC framework consists of two components: a structured neural network and a time-efficient basis algorithm. We retain the backbone of the low-complexity method while integrating the neural networks to calibrate the inputs. The existence of calibration mappings to improve system performance was proved in Ref. [99]. The NC employs a structured neural network architecture that leverages the permutation equivariance (PE) property inherent to wireless networks. PE is a property whereby an algorithm's output remains unchanged regardless of the order of its input elements. This phenomenon is common in wireless networks. Two representative PE properties are the uplink antenna-wise PE and downlink user-wise PE, where the order of antennas and users should not affect the algorithm's performance. By leveraging this property, NC-based algorithms can be generalized to different numbers of antennas and users [99]. This is important for terahertz UM-MIMO systems, whose design may involve high-dimensional problems with varying system scales, such as the number of users [85].

#### *4.1.3. Selecting basis algorithms*

Once the general algorithmic framework has been established, the next step is to select a (near) optimal basis algorithm tailored to a specific problem. This process requires expert knowledge of wireless systems. After identifying the basis algorithm, it is crucial to analyze its bottleneck modules and determine whether the bottleneck is due to the "hard-to-compute" problem, such as a large-scale matrix inversion, or the "hard-to-model" problem, such as the lack of prior distribution or optimal step size schedule.

Upon identifying the bottleneck module, it should be separated from the other components of the basis algorithm. Only the bottlenecks should be replaced with suitable neural networks. We analyze a few examples from existing research on model-driven DL for terahertz UM-MIMO.

- **Example 1:** The orthogonal AMP (OAMP) algorithm serves as a near-optimal basis algorithm to solve compressive channel estimation problems using FPNs. Each iteration consists of a linear estimator (LE) and a non-linear estimator (NLE). The LE of OAMP utilizes information from the system model to decouple the original compressive sensing problem into equivalent additive white Gaussian noise (AWGN) denoising problems for the NLE module to solve [124], and it can be derived in closed form. The bottleneck of the OAMP algorithm for channel estimation is the prior distribution required by the MMSE-optimal denoiser in the NLE module, which is challenging to obtain due to the complex channel characteristics of terahertz UM-MIMO. After identifying the NLE as the bottleneck module, the LE is kept unchanged in each iteration, and the NLE, which depends on the prior distribution, is replaced with a neural network [38].
- **Example 2:** The near-field multi-user beam-focusing problem presents "hard-to-compute" challenges due to the computational complexity of the weighted MMSE (WMMSE) algorithm in terahertz UM-MIMO [125]. Consequently, a low-complexity zero-forcing (ZF) scheme can be used as the basis algorithm. However, the bottleneck of ZF is its suboptimal performance when serving many users. Hence, we resort to the NC framework and employ neural networks to calibrate the input of the ZF beamformer, aiming to achieve performance comparable to that of iterative WMMSE algorithms but with lower complexity [85,99].

Hardware imperfections in the terahertz band mainly include mutual coupling, phase noise, antenna failures (e.g., gain and phase errors), and power amplifier (PA) non-linearity [18,74,126], among others. Currently, most studies on hardware-impaired terahertz systems are based on simulations, and different studies have considered different combinations. Hardware limitations can be grouped into three categories based on whether they influence the wireless channel $\boldsymbol{h}$, the measurement matrix $\boldsymbol{M}$, and the noise $\boldsymbol{n}$. Impairments affecting the channels, such as mutual coupling and antenna failures, can be absorbed into the channels, as the goal is usually to estimate the effective channel. Impairments affecting the measurement process, such as phase noise and PA non-linearity, can be considered in model-driven DL, as the LE is designed to decouple the measurements. However, these impairments can render the measurements a non-linear transformation of $\boldsymbol{y} = \boldsymbol{M}\boldsymbol{h} + \boldsymbol{n}$, where $\boldsymbol{y}$ denotes the received pilot signals. Hence, a more complicated LE should be utilized for decoupling. Basis algorithms compatible with non-linear the measurement process should be considered, such as those in Ref. [127]. Impairments affecting the noise term are often modeled as spatially correlated noise [18]. In these cases, the NLE of the model-driven DL is responsible for handling more complex noise types. If the noise statistics are known, whitening can be applied to simplify the problem, or the noise statistics can be incorporated into the neural network design to improve performance. Additionally, adversarial loss and test-time training [38] can be incorporated to enhance robustness to unexpected changes in noise statistics during online deployment.

#### *4.1.4. Loss function design*

Once the algorithmic framework and basis algorithm are determined, the next step is to design a loss function. The simplest approach is to use standard classification or regression losses based on relevant input-label pairs, referred to as direct loss functions. However, direct loss may be inefficient in some scenarios. This has motivated the development of improved indirect loss functions, including task-oriented and empirical Bayesian losses. Representative cases are discussed below.

- **Task-oriented losses:** Given the extremely large scale of terahertz UM-MIMO systems, obtaining optimal labels for standard loss functions is often computationally challenging or even impossible. Task-oriented losses can be utilized to tackle the "hard-to-compute" problem, especially for resource allocation problems. In an early work, Sun et al. [117] used the iterative WMMSE algorithm to create labels for training a neural network for interference management; however, this process may be prohibitively complex for large-scale systems. Instead, losses should be designed to bypass the expensive label generation step and directly train neural networks to solve the target task. These are referred to as task-oriented losses. One of the earliest examples of this approach is in Ref. [128]. For the sum-rate maximization problem in downlink multiuser beamforming, Huang et al. [128] proposed using the negative sum-rate directly as a task-oriented loss. This approach avoided the cumbersome label-generation process and achieved performance comparable to the WMMSE algorithm. Similar concepts have been widely applied in model-driven DL frameworks for terahertz UM-MIMO. For instance, Nguyen et al. [100] emphasized the importance of using an unsupervised task-oriented loss to directly minimize the objective function in hybrid beamforming design.
- **Empirical Bayesian losses:** Another important consideration in loss function design is the "hard-to-measure" problem. In terahertz UM-MIMO systems, obtaining a large channel dataset to train neural networks is difficult, particularly for channel estimation problems. This creates a chicken-and-egg problem: Accurate training data relies on channel estimation, but enhanced channel estimation is difficult without sufficient data. In channel estimation problems, the received pilot signals $\boldsymbol{y}$ are typically modeled as $\boldsymbol{y} = \boldsymbol{h} + \boldsymbol{n}$ or $\boldsymbol{y} = \boldsymbol{M}\boldsymbol{h} + \boldsymbol{n}$ [94]. In practice, only $\boldsymbol{y}$ are available, not the ground-truth $\boldsymbol{h}$. Therefore, the mean-square-error (MSE) loss adopted in many previous studies cannot be applied, that is, $\mathbb{E}\left\|\widehat{\boldsymbol{h}} - \boldsymbol{h}\right\|_2^2$, where $\widehat{\boldsymbol{h}}$ denotes the estimated channel and $\mathbb{E}$ denotes the expectation over wireless channels. To circumvent the dependence on ground-truth $\boldsymbol{h}$, empirical Bayesian methods may be adopted. For linear models, $\boldsymbol{y} = \boldsymbol{h} + \boldsymbol{n}$, Stein's unbiased risk estimator (SURE) is a good surrogate for the MSE loss [129]. Furthermore, for the generalized linear model $\boldsymbol{y} = \boldsymbol{M}\boldsymbol{h} + \boldsymbol{n}$, the generalized SURE (GSURE) can be adopted as an extension [130]. For high-dimensional problems in terahertz UM-MIMO, SURE and GSURE can be efficiently computed using only a few Monte Carlo trials, as reported in the literature, because of the high dimensionality of terahertz UM-MIMO channels [131]. Both SURE and GSURE have been successfully applied to UM-MIMO systems to achieve (near) MMSE optimal channel estimation performance in unknown environments [101,132]. The extension from white noise to spatially correlated Gaussian noise is discussed in Ref. [94], while recent results extend them to other exponential families of noise distributions [133], covering most noise types encountered in wireless communication. Empirical Bayesian losses facilitate unsupervised learning and adaptation in unknown environments without requiring prior knowledge or access to clean channels.

*4.1.5. Neural architecture design*

The neural architecture is also pivotal in enhancing the generalization and efficiency of model-driven DL. While the system and channel conditions in terahertz UM-MIMO systems may change rapidly owing to blockage, severe path attenuation, and hybrid-field effect, enhancing the generalization capability of the designed neural networks to accommodate these variations is crucial, including the number of users, far- and near-field multipaths, and SNR levels. Therefore, we introduce graph neural networks (GNNs) and hypernetworks as two candidate tools. Furthermore, because the limited number of multipaths in terahertz UM-MIMO channels suggests that the effective channel dimensions are significantly smaller than the number of antennas, it is feasible to work with a reduced-dimensional channel representation to lower the complexity of DL algorithms. Accordingly, we propose the neural RF radiance field (NeRF$^2$), Gaussian splatting (GS), and learning in the transform domain as potential tools.

- **Graph neural networks:** A promising direction for improving scalability and generalization is the design of neural architectures tailored to wireless networks. One important finding is that GNNs perfectly align with the permutation equivariance property of wireless networks [134], making them exceptionally effective in tackling large-scale resource management [135] and data detection problems [102,136]. Unlike traditional DL models, such as convolutional neural networks (CNNs) and MLPs, which often struggle with large-scale networks and new system settings, GNNs effectively utilize graph topology and the permutation invariance property inherent in wireless communications. In addition, in GNNs, the input–output dimensions of the neural network in each node are invariant with the number of users. This allows them to generalize across different system scales. For example, for the beamforming problem in an interference channel, a GNN trained on a small-scale network with 50 users can achieve near-optimal performance in a much larger-scale network with 1000 users [134]. Furthermore, GNNs are computationally efficient owing to their parallel execution capability and, to date, represent the only method capable of finding near-optimal beamformers for thousands of users within milliseconds [134]. This is particularly suitable for UM-MIMO networks with an extremely large system scale and a varying number of users. Please refer to Ref. [137] for a detailed overview of GNN applications in large-scale wireless networks.

- **Hyper-networks:** Owing to hybrid-field effects and blockage at terahertz frequencies, network parameters such as the number of far- and near-field paths, as well as SNR levels, can vary rapidly. When deploying AI-based algorithms in practical systems, a major challenge is ensuring that neural networks can adapt to such dynamics. Hypernetworks provide an effective solution [138]. A hypernetwork generates the weights for another neural network, known as the target network. Instead of fixed weights, the target network weights are dynamically produced by the hypernetwork based on its input. By setting parameters such as SNRs, number of paths, and speed of users as inputs to the hypernetwork, different weights can be generated for the target network under different system conditions. This allows the model-driven DL to seamlessly adapt to varying system conditions in the terahertz bands. Existing works in this direction include channel estimation [139], prediction [103], and joint source–channel coding [140].
- **Learning on the transform domain:** Owing to high path loss and limited diffraction in the terahertz band, terahertz UM-MIMO channels can be represented as a superposition of a limited number of paths. This characteristic implies a lower-dimensional representation of the terahertz channels compared to an ultra-massive number of antennas. Working on the transform domain of a channel with sparse features can reduce difficulty in training AI models for terahertz UM-MIMO channels. For channels where all multipaths lie in the far-field region, the array response vectors depend only on the angles of arrival, rendering the channel sparse in the angular domain. In such cases, the discrete Fourier transform (DFT) matrix is a suitable dictionary for sparse transformation. For near-field channels, Cui and Dai [141] first introduced an angle-distance domain dictionary, which was later enhanced with Newtonized dictionaries [90] and discrete prolate spheroidal sequence (DPSS) dictionaries [89] for improved performance. For generic hybrid-field channels, dictionary learning techniques can be used to obtain site-specific dictionaries [104]. A comparison between the results of Ref. [38] and Ref. [131] shows that applying learning in the sparse transform domain yields both improved performance and faster convergence for the same model-driven DL network (e.g., FPN-OAMP) used in channel estimation. Furthermore, learning in the transform domain can significantly reduce the number of training samples required to reach a performance level comparable to that achieved in the original non-sparse domain.
- **NeRF$^2$ and GS:** In a given propagation environment, once the positions of all transmitters are determined, the channel response at any position in that environment can be determined. Based on this principle, Google introduced the neural radiance field (NeRF) to model light field ray tracing. This concept was later extended to RF signals in Ref. [142], where NeRF$^2$ was proposed to model RF signal propagation as a continuous volumetric scene function. After training on limited measurements, NeRF$^2$ can provide accurate location-to-channel mapping, enabling prediction of the channel response at any spatial position within the environment. This enhances both the accuracy of channel modeling and the dimensionality reduction of high-dimensional wireless channels. Such mapping can be used in downstream tasks like channel estimation and prediction. Although the original experiments were conducted using a 5G massive MIMO array at sub-6 GHz frequencies, extending this idea to terahertz UM-MIMO systems holds considerable promise. However, a major limitation of NeRF$^2$ is its computational complexity [105]. In a typical setup, NeRF$^2$ requires approximately 200 ms to synthesize the channel characteristics for a given scene. This exceeds the requirements of most latency-sensitive applications. Wen et al. [105] recently proposed the utilization of 3D GS to accelerate the environmental aware channel modeling, called wireless radiation field GS (WRF-GS). Under the same setup, WRF-GS required only 5 ms for channel synthesis, which was significantly more efficient than NeRF$^2$.

*4.1.6. Case study 1: UM-MIMO beam-focusing*

We consider downlink beamforming (beam-focusing) [143] to maximize the sum rates in a multiuser UM-MIMO system operating in the near-field region [85]. We explain how to walk through the four essential steps to design an effective model-driven DL algorithm for this problem based on the NC framework.

- **Step 1: determining algorithmic frameworks.** In the first step, we determine the type of algorithmic framework to be applied. Although the WMMSE algorithm offers near-optimal results, its complexity makes it difficult to implement in UM-MIMO systems [125]. Therefore, we focus on non-iterative linear algorithms because of their lower computational complexity, and to enhance their performance using the NC framework to achieve a performance comparable to that of the WMMSE.
- **Step 2: selecting basis algorithms.** We select a ZF beamformer as the basis algorithm because of its good tradeoff between complexity and performance. The bottleneck of the ZF scheme is its suboptimality when the number of users is large. Hence, we designed a structured neural network module to calibrate the input to the ZF beamformer and enhance its performance, as shown in Fig. 5 [85].

- **Step 3: loss function design.** Using the direct loss function requires generating optimal labels, which requires repeatedly running high-complexity iterative WMMSE algorithms and is very expensive to implement. To avoid a cumbersome label generation process, we choose a task-oriented loss to directly minimize the negative sum rates of the considered system without labels. This significantly reduces the complexity of the training process.
- **Step 4: neural architecture design.** We design a structured neural network based on the PE property of wireless networks. We model the wireless network as a directed graph, with nodes representing the BS and users, and edges representing transmission links. The beam-focusing task is then framed as an optimization problem over this graph, where the PE property ensures that the order of user presentation does not affect system performance. Leveraging this property, we can deploy identical CNNs on every user, each sharing a common set of parameters, which dramatically lowers the total trainable weights and the training overhead. In addition, as identical CNNs use the same parameters, they can be applied to any number of users. Simulation results demonstrate that this approach can effectively generalize across a varying number of users.

Fig. 6 [85] shows the advantages of the proposed NC-based ZF beamformer. The performance upper bound is an iterative WMMSE algorithm with a significantly high computational complexity. The ZF and maximum ratio transmission (MRT) schemes were also compared. In the "matched" case, the number of users is the same during training and inference, while in the "mismatched" case, the training stage contains 75 users, but more users (i.e., 75–200) appear during inference. We highlight the following observations from the simulation results: First, it is shown that the proposed NC-based ZF outperforms both the original ZF and MRT schemes, and its performance compares well with the iterative WMMSE algorithm. In addition, the "mismatched" case indicates that the NC-based scheme trained on a small-scale network can be generalized to a much larger one. Finally, the proposed NC-based beamformer has significant advantages in terms of the computational complexity. Despite a similar performance, the runtime of the NC-based ZF scheme is only 0.16 s as compared to 19 s for the iterative WMMSE algorithm when the number of users is 150, which means that it is nearly 118 times more efficient. In this case study, we assumed a fully digital UM-MIMO system. It will be interesting to extend the NC framework further to the AoSA architecture in the future [144,145].

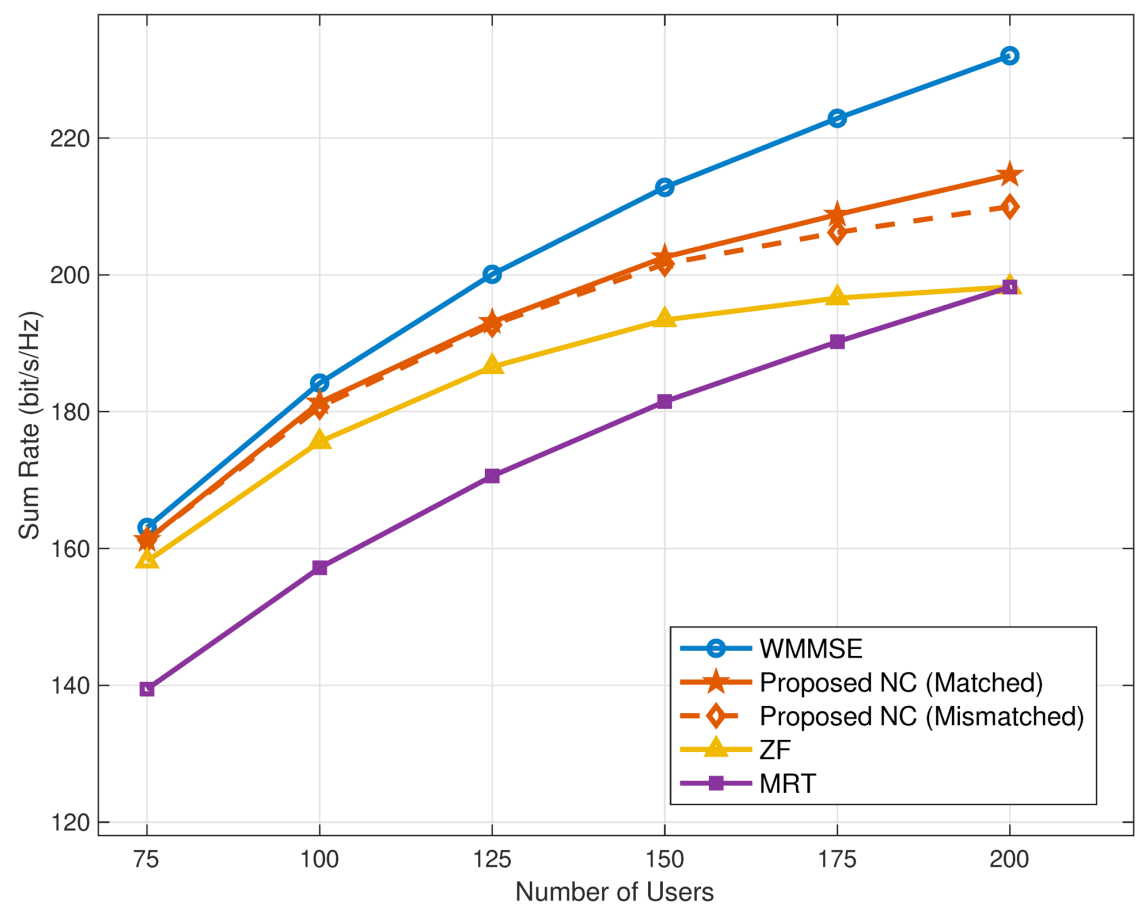

**Fig. 6.** Sum-rates as a function of the number of served near-field users. More detailed simulation settings can be found in Ref. [85]. We consider a 1024-element UM-MIMO array with −60 dBm noise power, 10 dB transmit power. MRT: maximum ratio transmission. Reproduced from [85] with permission.

*4.1.7. Case study 2: UM-MIMO data detection*

In the second case study, we switch our focus to the data detection problem, that is, detecting data symbols $\boldsymbol{x} = [x_1, x_2, \ldots, x_N]^{\mathrm{T}}$ from the received data signals $\boldsymbol{y}_{\mathrm{data}} = \boldsymbol{H}_{\mathrm{mul}}\boldsymbol{x} + \boldsymbol{n}$, where $N$ is the number of antennas, and $\boldsymbol{H}_{\mathrm{mul}}$ is the multiuser UM-MIMO channel. In UM-MIMO systems, the antenna array is large. In this case study, we demonstrate how model-driven DL can significantly enhance the data detection performance while maintaining low complexity [102,136].

- **Step 1: determining algorithmic frameworks.** In this case study, we work on iterative detection algorithms because of their better performance than closed-form linear detectors, and we employ to FPNs to enhance them.
- **Step 2: selecting basis algorithms.** Iterative data detectors consist of a linear module (LM) and a non-linear module (NLM) in each iteration. The OAMP [146] and expectation propagation (EP) [147] detectors can offer a near-optimal performance close to that of the maximum likelihood (ML) detector but entail a high-complexity matrix inversion operation in the LM that incurs cubic computational complexity with respect to the number of antennas. This is prohibitive in UM-MIMO systems with thousands of antennae. Conversely, AMP-based detectors are free of matrix inversion in LE, and their complexity is dominated only by the matrix-vector product [148]. However, AMP generally exhibits inferior performance compared with OAMP and EP. Given that computational complexity is a major bottleneck in UM-MIMO detection, we select the AMP detector as our basis algorithm and apply model-driven DL techniques to enhance its inversion-free LM and boost its performance to a near-optimal level with low complexity.

- **Step 3: loss function design.** We employ a direct loss function, which is the MSE loss between the detected and true symbols, to train the networks without special designs, like most previous works in this direction [102]. Efforts have been made to design improved loss functions based on the general framework of neural feature learning [149], which can train detectors to generalize to include different fading scenarios without the requirement of online training [150,151].
- **Step 4: neural architecture design.** The core of the LM in AMP-based detectors is the iterative decoupling of the posterior probability $p(\boldsymbol{x}|\boldsymbol{H}_{\text{mul}}, \boldsymbol{y}_{\text{data}})$ into a series of independent scalar probabilities $p(x_i|\boldsymbol{H}_{\text{mul}}, \boldsymbol{y}_{\text{data}})$, where $i = 1, 2, \ldots, N$ is the antenna index. In particular, $p(x_i|\boldsymbol{H}_{\text{mul}}, \boldsymbol{y}_{\text{data}})$ is assumed to be a Gaussian distribution independent of the other antenna indexes. The performance of the AMP detector is determined by the accuracy of this equivalent AWGN model, which is asymptotically true, but can be inaccurate when used with a finite number of antennas. Hence, in Ref. [102], we calibrated the LM of AMP using a neural network to learn an accurate approximation for the Kullback–Leibler divergence, enhance the accuracy of the AWGN model, and further improve the performance. Specifically, we chose GNNs because they obey the PE property of MIMO systems and can thus be trained more efficiently and generalized to different system scales.

In Fig. 7, we illustrate the symbol error rate (SER) performance of different detectors. As illustrated, the inversion-free AMP-GNN detector [102] significantly improves the performance of the AMP detector and its performance matches closely with that of the near-optimal EP detector with a high-complexity matrix inversion. In the “matched” case, the training and inference are both carried out in a $32 \times 24$ MIMO system. In contrast, in the mismatched case, we trained the AMP-GNN using a mixture of $32 \times 16$ and $32 \times 32$ channels and tested it on a $32 \times 24$ system. The mismatch in the system scale causes little performance loss for AMP-GNN, which verifies its good scalability.

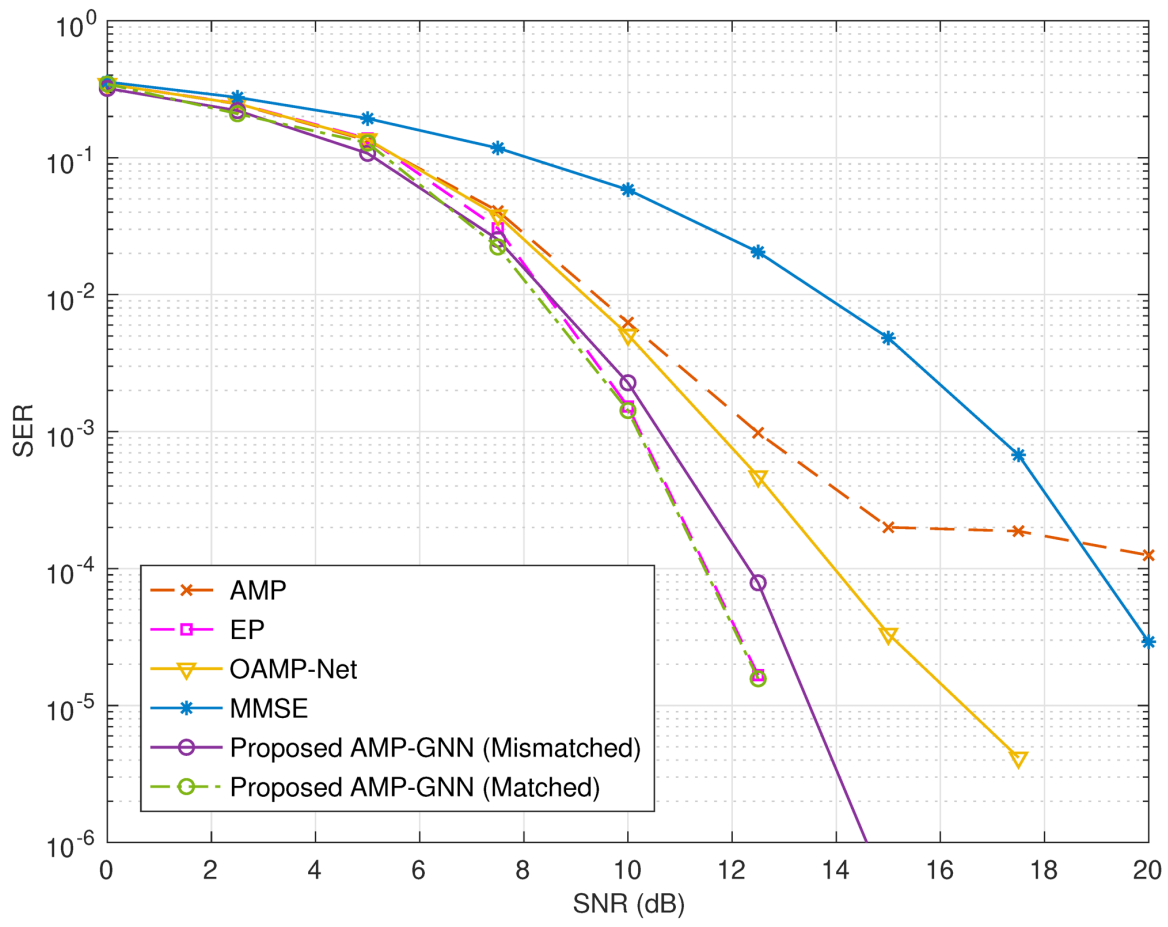

**Fig. 7.** SER comparison of AMP-GNN with other state-of-the-art detectors in a $32 \times 24$ MIMO system. More detailed simulation settings are presented in Ref. [102]. Since the number of RF chains is significantly smaller than that of antennas in terahertz UM-MIMO systems, baseband channels of $32 \times 24$ can correspond to systems with hundreds of antennas. Reproduced from Ref. [102] with permission.

*4.2. Roadmap 2: CSI foundation models*

In addition to model-driven DL, we propose a visionary concept known as the CSI foundation model. The central idea is that the designs of various transceiver modules share a common basis namely, the wireless channels. Knowledge of channel characteristics, such as distribution and second-order statistics, and features, such as sparsity and low rank, is fundamental to the design of all transceiver modules. Most existing studies train dedicated neural networks for different transceiver modules, such as channel estimation networks, CSI compression networks, and beamforming networks, each with its own task-oriented loss function. However, these separate networks can learn similar aspects related to the distribution of wireless channels. Training a dedicated neural network model for each transceiver module can cause significant redundancies in computational complexity, memory usage, and deployment costs. The existence of a common basis for the transceiver design, that is, a wireless channel, suggests the possibility of developing a unified foundation model. This model can provide essential information for designing a variety of downstream transceiver modules. This insight forms the basis for the CSI foundation models.

Specifically, the CSI foundation models aim to achieve two objectives. First, we train a neural network to learn the score function of wireless channels from limited channel data or directly from raw received signals. The inspiration for using the score function comes from the success of score-based generative models [152] and diffusion models [153]. A neural network that estimates the score function can serve as an open-ended prior for the design of various transceiver modules. Second, we develop model-driven DL frameworks that incorporate the physical layer foundation model as a flexible prior. This approach allows a single compact neural network model to function as a PnP prior, which enables the optimal design of many different transceiver modules.

We outline four essential steps for training and applying CSI foundation models and discuss an initial case study. A general research roadmap is shown in Fig. 4. The four steps include general frameworks, conditioning, site-specific adaptation, and joint design with model-driven DL. Finally, we present a case study that applies CSI-foundation models to UM-MIMO channel estimation based on Ref. [94].

Notably, a concurrent paper proposed a similar concept called "foundation models for wireless channels" [155], but with very different contents. Alikhani et al. [155] focused on learning representations and proposed a large-scale transformer-based pre-trained model for wireless channels to generate rich, contextualized embeddings that can outperform raw channels for downstream tasks. Conversely, our work focuses on the common foundations of wireless transceiver design and discusses the training and deployment of a single compact generative prior that can support a wide variety of downstream tasks in transceiver design. The perspectives of these two studies are complementary. Their intersection provides inspiration for the future development of CSI foundation models.

#### *4.2.1. General frameworks*

As shown in Fig. 4, the general framework comprises two levels. At the first level, we explore methods for training a neural network to estimate the score function of wireless channels. For (vectorized) wireless channels $\boldsymbol{h} \sim p(\boldsymbol{h})$, the prior score function is defined as the gradient of log-density function of the channels (i.e., $\nabla_{\boldsymbol{h}} \log p(\boldsymbol{h})$). We discuss the training strategies in two scenarios: ① when clean channel data are available, and ② when only raw received signals are accessible, which can be both noisy and incomplete (due to the limited RF chains in the AoSA). Next, we discuss how to combine the neural score function estimator with different sampling strategies to design various transceiver modules. We first discuss the level one, training strategies.

- **Score function from channel data:** When clean wireless channels $\boldsymbol{h}$ are available from measurement campaigns, we can train a neural network to approximate the score function, i.e., learning $\boldsymbol{s}_\theta(\boldsymbol{h}) \approx \nabla_{\boldsymbol{h}} \log p(\boldsymbol{h})$, where $\boldsymbol{s}_\theta(\boldsymbol{h})$ is the neural score function estimator with parameters $\theta$. We call such a network as "(prior) score network" for simplicity. The training of the score network has been extensively studied and can be achieved by denoising score matching [156–158]. If clean wireless channels are available, we can directly pre-train based on these methods to obtain a score network. In addition, as the effective dimensionality of high-frequency wireless channels is small because of the limited number of paths, it may be beneficial to first transform the channel into an appropriate sparse domain and then perform training on such a domain, which is similar to the idea behind latent diffusion models [159].
- **Score function from raw received signals:** Depending on the system architecture, the received pilot signals can be either $\boldsymbol{y} = \boldsymbol{h} + \boldsymbol{n}$ in fully digital systems [160] or $\boldsymbol{y} = \boldsymbol{M}\boldsymbol{h} + \boldsymbol{n}$ in hybrid analog-digital systems (such as the AoSA architecture) [3,34]. In the former case, the (prior) score function $\nabla_{\boldsymbol{h}} \log p(\boldsymbol{h})$ can be estimated using empirical Bayesian methods based on the SURE loss [129,161]. Meanwhile, for the AoSA architecture, the GSURE loss [130] can be adopted as an extension.

We then discuss the second level, that is, how to apply CSI foundation models as open-ended priors to enable a variety of downstream tasks in the transceiver design. We discuss four cases in the following paragraphs, including prior sampling, posterior sampling, sequential sampling, and joint sampling, which are certainly non-exhaustive. However, further interesting applications are yet to be discovered.

- **Prior sampling for data augmentation:** Once the score network $\boldsymbol{s}_\theta(\boldsymbol{h})$ has been pre-trained on the available channel data or the raw received signals, we can then sample from it by using score-based generative models to synthesize more channel data [152]. Synthetic channel data can be combined with measured channel data for data augmentation to create an expanded dataset that supports various downstream tasks in the transceiver design. Initial results have demonstrated the effectiveness of score-based data augmentation in enhancing the performance of DL-based channel compression, channel coding, uplink–downlink channel mapping, beam alignment, hardware impairment mitigation, constellation shaping, and wireless fidelity (Wi-Fi) sensing [91,92,162–164].

- **Posterior sampling for inverse problems:** Inverse problems, including channel estimation and tracking, are common in wireless transceiver design. For terahertz UM-MIMO systems, the system model for channel estimation can be simplified as $\boldsymbol{y} = \boldsymbol{M}\boldsymbol{h} + \boldsymbol{n}$ after vectorization. In particular, $\boldsymbol{M}$ is the measurement matrix jointly determined by the pilot symbols and pilot combiners, and is perfectly known in the system, which is a fat matrix because the number of RF chains is significantly less than that of the antennas in the AoSA architecture. We are interested in estimating the wireless channel $\boldsymbol{h}$ from the received pilot signals $\boldsymbol{y}$. Suppose that the channels follow a prior distribution $\boldsymbol{h} \sim p(\boldsymbol{h})$, the channel estimation problem can be solved from a Bayesian perspective by posterior sampling based on diffusion models. The forward diffusion process follows a Markov chain with gradually added Gaussian noise from $\boldsymbol{h}_0$ to $\boldsymbol{h}_T$, where $\boldsymbol{h}_t$ denotes the generated channels at time step $t$ $(t = 0,1,2,\ldots,T)$, and $T$ is the maximum number of time steps. The reverse sampling process follows from $\boldsymbol{h}_T$ to $\boldsymbol{h}_0$. To incorporate information from the system model, the posterior score is given by $\nabla_{\boldsymbol{h}_t} \log p(\boldsymbol{h}_t|\boldsymbol{y}) = \nabla_{\boldsymbol{h}_t} \log p(\boldsymbol{h}_t) + \nabla_{\boldsymbol{h}_t} \log p(\boldsymbol{y}|\boldsymbol{h}_t)$ by using the Bayes' rule. In this equation, the prior score $\nabla_{\boldsymbol{h}_t} \log p(\boldsymbol{h}_t)$ is available from pre-trained CSI foundation models. We need to derive the likelihood score $\nabla_{\boldsymbol{h}_t} \log p(\boldsymbol{y}|\boldsymbol{h}_t)$ based on the system model $\boldsymbol{y} = \boldsymbol{M}\boldsymbol{h} + \boldsymbol{n}$ to complete the picture. Meng and Kabashima [165] proposed a technique called noise-perturbed pseudo-likelihood score to approximate the true likelihood score with low computational complexity for the linear system model $\boldsymbol{y} = \boldsymbol{M}\boldsymbol{h} + \boldsymbol{n}$. They further extend a similar idea to quantized inverse problems $\boldsymbol{y} = Q(\boldsymbol{M}\boldsymbol{h} + \boldsymbol{n})$ in Ref. [106], where $Q(\cdot)$ denotes a uniform quantizer. This technique has recently been applied to channel estimation in quantized MIMO systems with low-resolution analog-to-digital converters (ADCs) [166].
- **Sequential sampling for compression:** CSI compression and feedback are fundamental components of frequency-division duplex (FDD) massive MIMO systems. In the downlink of such systems, users must estimate the channels and feed them back to the BSs. Previous compression algorithms are mostly based on discriminative rather than generative learning [120,167]. Generative CSI foundation models can also be used for the compression and decompression of CSI. The basic idea is sequential sampling, which uses a posterior sampler based on pre-trained CSI foundation models to iteratively gather informative channel measurements for compression via adaptive compressed sensing [107], and then decompress the codewords by reversing these steps to reconstruct the channels.
- **Joint sampling for cross-module design:** In UM-MIMO systems, the pilot overhead required for accurate channel estimation is significant owing to the high dimensionality and limited number of RF chains. To address this issue, a joint channel estimation and data detection algorithm can be considered to reduce pilot overhead. Such a cross-module design can also be easily supported by CSI foundation models. To accomplish this goal, one can first establish a diffusion process that represents the joint distribution of the channels and symbols given the noisy received pilot signals and subsequently run the reverse denoising process to generate samples. In contrast to the continuous prior distribution of the channels, the prior distribution of the symbols is discrete. Thus, computing the score function of the symbols in the diffusion model is difficult. To address this, Zilberstein et al. [108] and Zilberstein et al. [168] proposed an annealed Langevin dynamics algorithm to incorporate the discrete nature of the constellation elements so that joint sampling can work. In future research, it would also be interesting to discover other possible cross-module designs, such as joint channel estimation and decoding.

*4.2.2. Conditioning*

In practical applications, it is important to enhance the generality of CSI foundation models for effective adaptation to various scenarios and environmental conditions without retraining. Conditional generation plays a crucial role in achieving this objective by incorporating a conditional label, denoted as $\boldsymbol{c}$, into the score function, such as $\boldsymbol{s}_\theta(\boldsymbol{h}, \boldsymbol{c})$. This label specifies the particular type of the generated channel, enabling the model to tailor its output to specific conditions [169,170].

Several advantages can be harnessed using the conditioning technique for CSI foundation models. First, it enhances efficiency. Using conditioning, a single model can generate diverse types of channels by leveraging conditional generation, which eliminates the need to train multiple models and facilitates practical deployment. For example, when the conditional label is weather, the same model can serve as a score network and generate channels for sunny and rainy days. In addition, this increases generalization capability. Well-trained conditional models can generate a wide variety of channel samples with different characteristics, which enhances the capability of the model to generalize beyond the original training data and improves data augmentation performance. Commonly used conditional labels include the position in the environment and link status, such as uplink/downlink and LoS/non-LoS. Additionally, incorporating multimodal information into conditional labels is a promising direction. For instance, data from vision cameras can effectively assess the environmental status and serve as a potential conditional label. However, incorporating multimodal data for conditioning can be complex and expensive. It is advisable to borrow ideas from semantic [171] or task-oriented communication [172,173] to generate labels that capture the minimal sufficient information of the environment.

#### 4.2.2. Site-specific adaptation

When deploying CSI foundation models at different sites, performance can degrade because of environmental differences. Therefore, it is important to design site-specific adaptation schemes to fine-tune the models. The vanilla scheme fine-tunes the entire model in a new environment. However, this approach can be prohibitive when the number of parameters is high. Considering the potential scale of the foundation model, it is important to design parameter-efficient fine-tuning (PEFT) schemes that retrain only a small portion of all parameters to adapt to the new environment. We may resort to existing fine-tuning schemes for LLMs for inspiration. The adapter tuning scheme adds a few linear layers inside the transformer for better results but can increase the inference delay [174]. Bit-fit methods only fine-tune the bias terms, which leads to a lower computational cost but degrades performance [175]. One of the most popular methods is the low-rank-adaptation (LoRA) scheme [176]. LoRA is a fine-tuning technique for transformer models that freezes pre-trained weights and introduces trainable low-rank matrices into each layer. This reduces computational complexity and preserves parallel processing capabilities. When multiple devices (such as BSs) participate in the fine-tuning of CSI foundation models, LoRA can be further extended to a federated LoRA over wireless networks to support collaborative PEFT [109,177] and simultaneously protect the privacy of raw data [178,179].

#### 4.2.4. Joint design with model-driven DL

On the one hand, the sampling strategies mentioned above, combined with the learned score network, already belong to model-driven DL frameworks. However, the score function is related to the MMSE denoiser for various types of exponential noise families. Please refer to Ref. [133] for a detailed description. Hence, it is natural to integrate it with AMP-family algorithms [180] through the denoising AMP framework [124] to solve a variety of inverse problems in physical layer communications. The idea of "separation" is the core of the joint design of model-driven DL and CSI foundation models. Model-driven DL separates system-specific attributes (e.g., those typical of a certain system model) from system-agnostic characteristics shared across different system architectures and transceiver modules (e.g., the prior distribution of channels). Model-driven DL provides an interface for separation and considers system-specific factors. As a complement, CSI foundation models can provide prior knowledge regarding shared system-agnostic components and can be seamlessly combined with different model-driven DL frameworks to solve diverse problems.

#### 4.2.5. Case study: CSI Foundation Models

We present an initial case study on training the CSI foundation model using denoising score matching with the raw pilot signals received and then apply it to UM-MIMO channel estimation. The simulation details are available in Ref. [94].

- **Step 1: determining general framework.** We consider training the score network from the raw received signals, $\boldsymbol{y} = \boldsymbol{h} + \boldsymbol{n}$. This corresponds to a fully digitalized system model. For the AoSA architecture in UM-MIMO systems, such a system model can be achieved by stacking the received signals from multiple timeslots. Slightly different from the original physical layer foundation model described above, we train the score network $\boldsymbol{s}_{\theta}(\boldsymbol{h})$ by using denoising auto-encoders based on $\boldsymbol{y}$ [94]. Based on the score network and Tweedie's formula [181], the MMSE-optimal channel denoiser can be obtained in closed form for fully digital systems, i.e., $\boldsymbol{y} = \boldsymbol{h} + \boldsymbol{n}$. For the general hybrid analog-digital system model $\boldsymbol{y} = \boldsymbol{M}\boldsymbol{h} + \boldsymbol{n}$, we chose to incorporate the denoiser as the NLE of the OAMP algorithm to jointly design it with the model-driven DL.
- **Step 2: conditioning.** In this initial case study, we did not consider conditioning and pretrained an unconditional score network, owing to limited time. Nevertheless, it is easy to follow the conditioning procedure introduced earlier to further enhance the model.
- **Step 3: site-specific adaptation.** Since here we train the score network $\boldsymbol{s}_{\theta}(\boldsymbol{h})$ based on the received pilot signals $\boldsymbol{y}$, the denoising score matching loss can be computed solely based on $\boldsymbol{y}$. Hence, the model can be adapted whenever there are new received pilot signals arriving. We used Vanilla online learning to tune the model and found that it could quickly adapt to changes in the channel covariance matrix within a couple of pilot transmissions [94].
- **Step 4: joint design with model-driven DL.** The AoSA architecture results in fewer RF chains than antennas. Consequently, UM-MIMO channel estimation becomes a compressive sensing problem. We adopt the FPN framework and select OAMP as the basis algorithm. The LE of the OAMP can be derived in a closed form. It utilizes information from the system model to decouple the original compressive sensing problem into equivalent AWGN denoising problems for the NLE module to solve. The bottleneck of the OAMP lies in its unknown prior distribution. Bayes-optimal OAMP requires the NLE to be the MMSE channel denoiser [124,146], which, luckily, can be derived based on the physical layer foundation model by using Tweedie's formula [94,181].

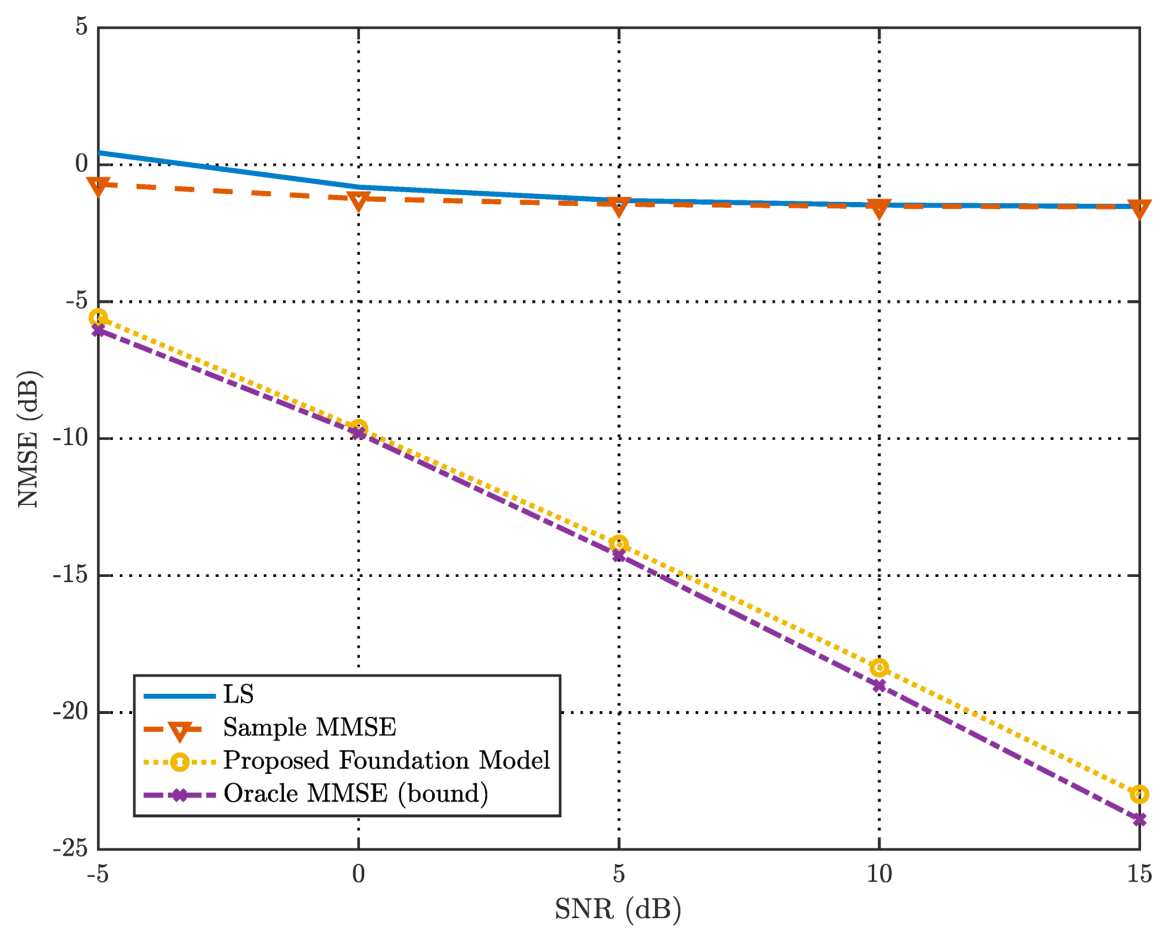

**Fig. 8.** NMSE versus the received SNR when the under-sampling ratio is 0.3. More detailed simulation settings can be found in Ref. [94]. We consider a 1024-antenna UM-MIMO array. The under-sampling ratio is the number of columns in measurement matrix $\boldsymbol{M}$ divided by its number of rows. Reproduced from Ref. [94] with permission.

In Fig. 8 [94], we present the normalized mean square error (NMSE) performance of compressive channel estimation in UM-MIMO systems as a function of the received SNR. We adopted a spherical wave model for the near-field UM-MIMO channel. The performance bound is the oracle MMSE method, which assumes that perfect knowledge of the channel distribution is available. As observed, the proposed physical layer foundation model performs close to the bound across different SNR levels and significantly outperforms both the LS and sample MMSE estimators, verifying its effectiveness. In this case study, we did not consider the wideband beam squint effect, and the proposed framework can be readily extended to handle such scenarios in a manner similar to that in Refs. [38,182].

### *4.3. Roadmap 3: Applications of LLMs*

Recent breakthroughs in LLMs, such as GPT-4 [183] and DeepSeek-R1 [184], have transcended the original scope of natural language processing, demonstrating transformative capabilities in diverse fields, including code synthesis, scientific discovery, and autonomous system control [185]. Although LLMs have the potential to reshape wireless network designs, their applications in terahertz systems remain unexplored. To the best of our knowledge, no previous study has explicitly investigated LLMs in terahertz systems. This section summarizes existing works in the general area of LLM-enabled wireless networks and analyzes how they may be adapted to systems with terahertz-specific challenges. The discussion is divided into five subsections: LLMs for estimation, optimization, searching, network management, and protocol understanding. Because this is an emerging direction, the ideas and methods discussed here are certainly not exhaustive. In Section 5, we discuss future research directions and open issues for integrating LLMs with terahertz systems.

#### *4.3.1. LLMs for estimation*

Estimation problems, including channel estimation, prediction, tracking, localization, and sensing, play pivotal roles in physical-layer communication. We summarize the possible ways in which LLMs may play a role in these areas.

- **LLMs as backbones:** LLMs have demonstrated powerful capabilities in cross-modal tasks. Liu et al. [110] proposed to freeze most parameters of a pretrained GPT-2 model and fine-tuning only a limited number of parameters for cross-modal knowledge transfer from the feature space of LLMs to the wireless channels. They then applied the fine-tuned model to the uplink-to-downlink channel prediction problem and achieved an improved performance. In Ref. [186], the model was further extended to handle general space-time-frequency wireless channels. By pre-training over a massive channel dataset, the model can be generalized for channel prediction in various CSI configurations. Yu et al. [187] incorporated multimodal data and proposed a ChannelGPT model for wireless networks. In Refs. [188–190], fine-tuning LLM models are proposed to serve multiple physical layer tasks, leveraging the multitask learning capabilities of large AI models.
- **LLMs as hyper-networks:** Zhang et al. [191] investigated the use of LLMs in the context of the least absolute shrinkage and selection operator (LASSO) problem. Although their focus was not directly on wireless systems, many estimation problems in wireless systems can be similarly formulated as sparse reconstruction tasks, making this study highly relevant. The proposed LLM-LASSO framework introduces a novel paradigm in which pretrained LLMs dynamically generate context-aware regularization parameters or feature selection masks by interpreting problem-specific metadata like hypernetworks. LASSO estimators are particularly well suited for terahertz UM-MIMO channels because they can be represented as high-dimensional sparse vectors. It is valuable to investigate how contextual information regarding wireless propagation can enhance the regulation parameter generation of terahertz UM-MIMO channel estimators.

#### *4.3.2. LLMs for optimization*

Recent advancements in LLMs have facilitated their integration into optimization tasks, particularly in wireless communication systems. LLMs, which leverage their pre-trained reasoning capabilities and multimodal data processing, can offer a paradigm shift from conventional optimization frameworks. Unlike traditional methods which rely on delicate problem formulation and modeling by human experts and training task-specific DL models, LLMs can automatically interpret and formulate the problem from language-based and possibly multimodal descriptions and provide zero- or few-shot solutions to different optimization problems, such as power control [192,193], traffic prediction [193], and spectrum sensing [111], and so on, without further retraining. Iterative prompting and chain-of-thought allow LLMs to refine their optimization decisions through feedback information. In addition, LLMs can be combined with model-driven DL-based optimization algorithms. They can play a role similar to that of hypernetworks in generating context-specific network parameters based on multimodal descriptions of the environment. Nazar et al. [194] proposed ENWAR, an LLM framework for multisensory data that employs retrieval-augmented generation for the real-time environmental perception of wireless networks. It can serve as a feature extractor to work hand in hand with LLM-based optimization algorithms. Owing to the channel characteristics of terahertz UM-MIMO systems, radio signals are easily subjected to blockage and misalignment. LLMs can utilize multimodal sensory data to enhance situational awareness and make optimization decisions based on their own intelligence without the aid of human experts.

#### *4.3.3. LLMs for searching*

Compared with deploying LLMs online for estimation and optimization problems, which must consider computational and memory complexity issues, it seems more practical to leverage LLMs to solve offline problems. Examples include large-scale wireless network planning (e.g., searching for the optimal antenna downtilt angles), facility localization (e.g., searching for the optimal number and positions of BSs), and sparse array design (e.g., searching for the optimal antenna position for a given aperture and number of antennas) in the context of terahertz UM-MIMO systems. These problems have several common characteristics. First, they are large-scale combinatorial problems involving integer variables, making them particularly difficult to solve as the system dimensions increase. Therefore, efficient heuristic-search algorithms are required. Second, these problems are typically solved once and for all. When an optimal solution is obtained offline, continuous updates are not required.

For such offline tasks, LLMs are especially promising because of their extensive knowledge base and ability to inspire the design of novel heuristics. For instance, a FunSearch paper from Google illustrates how LLMs can facilitate mathematical discoveries by navigating several candidate algorithms through a program search to uncover effective heuristics that can exceed the performance of the best human experts [112]. A similar methodology was adopted for challenges in communication systems, in a recent study on the LLM-guided search for deletion-correcting codes, where LLMs assist in identifying efficient error-correcting codes [195]. Considering the scale of terahertz UM-MIMO systems, efficient heuristics are extremely important. Similarly, by harnessing the capabilities of LLMs, the efficiency of combinatorial search problems that are prevalent in the deployment of terahertz-band wireless systems can be enhanced.

#### *4.3.4. LLMs for network management*

Recent advances in LLM agents have underscored a transformative trend toward autonomous and intelligent network management. For instance, Shen et al. [113] introduced an autonomous edge AI system that considered a client-edge-cloud hierarchical architecture. A GPT-based LLM is deployed in the cloud to understand natural language inputs and then plan and generate code. The system is designed to dynamically coordinate distributed AI models at the network edge to fulfill diverse user requirements, such as triggering federated learning tasks to update the models, thus demonstrating the capability of autonomously managing edge AI systems. Tong et al. [196] proposed the WirelessAgent framework, in which LLMs are leveraged as AI agents to facilitate intelligent network slicing in 6G. LLM agents are designed to process multimodal data and use techniques such as retrieval-augmented generation to facilitate intent understanding and decision making. In addition to the development of LLMs for network management, transformative innovations in network architecture are required to support the deployment of large AI models. Researchers at China Telecom proposed a new concept called “AI flow,” which introduced a novel framework designed to efficiently deploy multimodal LLMs across heterogeneous network infrastructures [197]. They proposed a dynamic distribution of inference tasks based on the available resources, network conditions, and task requirements. In terahertz systems, the substantial bandwidth enables high-rate and low-latency wireless communications capable of supporting LLMs with extensive data throughput. Nevertheless, the blockages and dynamic propagation environments prevalent in the terahertz band introduce additional uncertainty. The exploration of LLM-based autonomous network management is important for intelligent interference management and handover control. Investigating autonomous link adaptation and network slicing in response to the diverse computational demands of AI inference tasks is also meaningful.

*4.3.5. LLMs for protocol understanding*

Recent efforts have also demonstrated the significant potential of LLMs for understanding complicated communication protocols and standard documents. Nikbakht et al. [198] provided a comprehensive dataset spanning 3rd Generation Partnership Project (3GPP) releases from Releases 8 to 19, which enabled the fine-tuning of LLMs for questions and reasoning tasks in the telecom domain. Zou et al. [114] and Bariah et al. [199] illustrated the effectiveness of fine-tuning pre-trained LLMs for accurately classifying and extracting information from 3GPP technical documents. In addition, Bornea et al. [200] and Yilma et al. [201] leveraged retrieval-augmented generation by integrating telecom-specific knowledge bases into an LLM processing pipeline. Collectively, these studies highlight a growing body of literature that improves the understanding of protocols in telecommunications. In the future, it will be meaningful to develop domain-specific LLMs to understand the protocols in terahertz-band wireless systems.

## 5. Future directions and open issues

After discussing the three research roadmaps, we offer the following takeaway messages covering lessons learned, future directions, and open issues.

*5.1. Remarks on roadmap 1: Model-Driven DL*

The essence of model-driven DL is to leverage the expert knowledge inherent in the system and allow AI to focus only on bottleneck modules that are either “hard-to-compute” or “hard-to-model.” It is important not to “reinvent the wheel” by attempting to learn something that is already available in the system model. It is preferable to design appropriate loss functions and neural architectures tailored to the specific features and requirements of the problem. Always keeping the role of the “model” in mind is fundamental to the success of model-driven DL.

For future directions, on the one hand, it is important to keep tracking the development of signal processing and optimization algorithms and leverage state-of-the-art techniques as the basis for model-driven DL. This requires a deep understanding of both the specific problem at hand and in-depth knowledge of classical model-based algorithms. Furthermore, the computational and hardware costs associated with the deployment of model-driven DL must be carefully considered. It is important to design hardware-friendly algorithms considering the scale of terahertz UM-MIMO systems.

However, it is important to strengthen the connections between model-driven DL and CSI foundation models. Model-driven DL incorporates algorithm-specific designs to improve the convergence and performance of traditional algorithms. As mentioned previously, many components of transceiver algorithms share a common foundation, the wireless channel, which highlights the potential for connecting model-driven DL with CSI foundation models. Such an approach would allow a shared knowledge base to be adapted for specific tasks, thereby enhancing the overall performance. Furthermore, establishing these connections is crucial for advancing the integration of theoretical insights and practical applications in wireless communication.

*5.2. Remarks on roadmap 2: CSI Foundation Models*

The essence of CSI foundation models is to identify and separate the common ground of diverse tasks and focus on learning these “foundations.” The goal is to enable a single compact foundation model to contribute to the design of a wide variety of transceiver modules. A few frameworks may be instrumental in separating the foundations of different problems. One such example is Bayesian inference, where the prior and likelihood can be separated [202]. Denoising-based AMP and posterior sampling are two important examples in this category. Both are related to the denoiser of the wireless channels under Gaussian noise, which is further connected to the score function of the wireless channels. Therefore, tracking the latest developments in signal denoising and score estimation is important. Another example is the information-theoretic framework of neural dependence decomposition, which separates feature learning from feature usages [149], and learns universal features for various downstream tasks. It is worthwhile to follow these directions and dig deeper into the core of the CSI foundation models while discovering more applications in this promising field.

In addition, it is crucial to investigate the neural network architectures underpinning CSI foundation models. Although not discussed in detail here, it is meaningful to draw insights from the advanced neural architectures developed in the broader machine learning community to strengthen our understanding of these models. It is also worthwhile to keep an eye on the state-of-the-art developments of LLMs, as scaling CSI foundation models for real-world applications, such as terahertz UM-MIMO systems with multiple BSs and a large number of users, will likely encounter similar scalability and computational challenges. Therefore, learning from the evolution of LLM architectures could be instrumental in the successful large-scale deployment of CSI foundation models.

### *5.3. Remarks on roadmap 3: Applications of LLMs*

Despite the empirical performance gains provided by LLMs, the underlying reasons why pretrained LLMs achieve success in wireless tasks remain unclear. Moreover, a critical open question is whether the inherent complexity of LLMs is disproportionate for wireless channels, given their highly structured and often low-dimensional characteristics. Furthermore, the cost of deploying LLMs in wireless systems remains a challenge, necessitating rigorous analysis of their memory efficiency, energy efficiency, sample complexity, and inference latency. When it comes to terahertz UM-MIMO systems, due to the enormous scale of the system, we must be more careful about these practical concerns, especially the “hard-to-compute” problems mentioned earlier. Finally, LLMs require substantial data for both training and fine-tuning, which often necessitates extensive measurement efforts and poses “hard-to-measure” challenges. Additionally, some telecommunication data are subject to privacy concerns. One promising research direction is to explore fine-tuning telecom-specific LLMs in wireless digital twins, where data are abundant, before deploying them in real-world environments with minimal tuning. Another promising approach is to investigate distributed or federated fine-tuning, which enables cooperative development across many telecom users, while respecting data privacy.

## 6. Conclusions

In this paper, we presented a methodological study on the application of AI to address the complex challenges in terahertz UM-MIMO systems. Although both AI and terahertz technologies are recognized as crucial enablers of 6G and beyond systems, their intersection remains in its early stages. Therefore, this study aimed to bridge this gap. The first half of this paper focuses on the system and channel characteristics of terahertz UM-MIMO systems, illustrating how these challenges have naturally prompted the application of AI. The second half outlines three research roadmaps–model-driven DL, CSI foundation models, and applications of LLMs–which are promising directions for developing AI solutions for terahertz UM-MIMO systems. We outlined the essential steps of these roadmaps, combining high-level concepts with detailed explanations, and analyzed several representative case studies. Substantial progress has been made in this emerging area of interdisciplinary research. Finally, we presented our vision for future research and identified key open issues.

## Acknowledgments

This work was supported in part by the Hong Kong Research Grant Council (16209023).

## Compliance with ethics guidelines

Wentao Yu, Hengtao He, Shenghui Song, Jun Zhang, Linglong Dai, Lizhong Zheng, and Khaled B. Letaief declare that they have no conflict of interest or financial conflicts to disclose.